\documentclass[aps,prl,amsmath,amssymb,reprint,notitlepage]{revtex4-1}
\pdfoutput=1
\usepackage{graphicx}
\usepackage{dcolumn}
\usepackage{bm}
\usepackage{tabularx}
\newcolumntype{C}{>{\centering\arraybackslash}X}
\usepackage{multirow}
\usepackage[font={footnotesize}]{caption}
\usepackage{color}
\usepackage{amsmath}
\usepackage{braket}
\usepackage{units}

\DeclareMathOperator{\1i}{i}
\usepackage[hidelinks]{hyperref}
\hypersetup{
     colorlinks   = true,
     citecolor    = blue,
     linkcolor    = red,
     urlcolor     = black
     }

\begin{document}

\title{Two-colour interferometry and switching through optomechanical dark mode excitation}

\author{David P. Lake}%

\author{Matthew Mitchell}

\author{Barry C. Sanders}

\author{Paul. E. Barclay}
\email{pbarclay@ucalgary.ca}

\affiliation{Department of Physics and Astronomy and Institute for Quantum Science and Technology, University of Calgary, Calgary, AB, T2N 1N4, Canada}

\begin{abstract}
Efficient switching and routing of photons of different wavelengths is a requirement for realizing a quantum internet. Multimode optomechanical systems can solve this technological challenge and enable studies of fundamental science involving widely separated wavelengths that are inaccessible to single-mode optomechanical systems. To this end, we demonstrate interference between two optomechanically induced transparency processes in a diamond on-chip cavity. This system allows us to directly observe the dynamics of an optomechanical dark mode that interferes photons at different wavelengths via their mutual coupling to a common mechanical resonance. This dark mode does not transfer energy to the dissipative mechanical reservoir and is predicted to enable quantum information processing applications that are insensitive to mechanical decoherence. Control of the dark mode is also utilized to demonstrate all-optical, two-colour switching and interference with light separated by over 5 THz in frequency.
\end{abstract}

\maketitle

\section*{Introduction}
\noindent Interference is a ubiquitous physical phenomenon central to applications ranging from detection of gravitational waves \cite{ref:LIGO2017gw170104} to the implementation of modulators and other integrated photonics technology \cite{ ref:yariv2006poe, ref:liu2004hss,ref:obrien2009pqt}. Each of these examples interfere light at, or near, the same wavelength to convert differences in phase to changes in intensity. The emergence of frequency bin qubits \cite{ref:clemmen2016ramsey, ref:lu2018quantum, ref:kues2017chip} and a desire to interface quantum networking components based on different photonic technologies \cite{ref:kurizki2015quantum} has created the need for devices that interfere light with widely separated frequencies. Typically, this challenge has been addressed using nonlinear atomic \cite{ref:harris1998photon, ref:chang2014quantum} or solid-state \cite{ref:mcguinness2010quantum, ref:kobayashi2016frequency, ref:li2019tunable, ref:Qu2019CBD} materials, whose nonlinear optical susceptibility combined with precise photonic dispersion engineering can mediate interactions between different wavelengths of light. Here we demonstrate that cavity optomechanics \cite{ref:aspelmeyer2014co} provides a realization of multi-colour optical interference that can be implemented in transparent linear materials through light's interaction with a nanofabricated mechanical resonator. This process does not suffer from saturation given the simple harmonic oscillator nature of our excitations, and can be realized with effective cooperativity exceeding unity.

By coherently coupling light confined in an optical cavity to the motion of a mechanical resonance of the same cavity, light can be slowed and stored \cite{ref:safavi2011eit, ref:weis2010oit, ref:fiore2013ops}. The signature of this coherent optomechanical coupling between phonons and photons is a narrow transmission window in the otherwise opaque optical cavity resonance spectrum, referred to as optomechanically induced transparency (OMIT) \cite{ref:weis2010oit, ref:safavi2011eit}, whose highly dispersive optical response is independent of the phase of its input field. However, phase can be critically important to the optical properties of multimode optomechanical systems, in which multiple optical fields are injected into an optomechanical device. In the optical domain, multimode cavity optomechanical devices have been used for experimental demonstrations of wavelength conversion \cite{ref:hill2012cow, ref:dong2012ODM, ref:liu2013eit}. Microwave-frequency multimode devices have enabled low-noise frequency conversion \cite{ref:ockeloen2016LNA} and entanglement between photons \cite{ref:barzanjeh2018ser}, whereas hybrid electro-optomechanical devices have used coherent interference between optomechanically and piezomechanically driven motion to bridge microwave and optical frequencies \cite{ref:balram2016ccr}. Optical-frequency optomechanical devices whose intensity response is sensitive to the relative phase of multiple input optical fields with widely separated wavelengths, i.e., that involve interference between different colours of light, have yet to be reported despite proof-of-principle demonstrations based on atomic media \cite{ref:korsunsky1999phase,ref:kang2006pcl,ref:kang2011pcs, ref:Kim2015ALS} and recent advances in cavity optomechanical mediated coupling between multiple mechanical modes \cite{ref:xu2016topological, ref:Kuzyk2017MultiInt}.

\begin{figure}[ht]
\centering
\includegraphics[width=\linewidth]{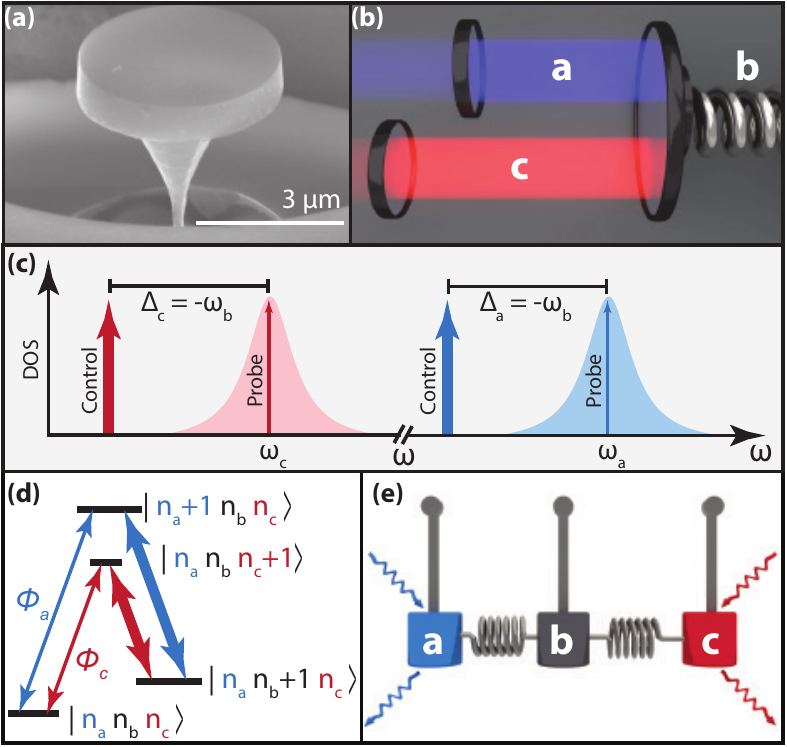}
\caption{Multimode optomechanical system coupling two optical modes to a mechanical mode of a diamond microdisk. (a) Scanning electron micrograph of a diamond microdisk similar to the $\sim\!5\ \mu$m diameter one studied in this work. (b) Schematic of a multimode optomechanical system where two Fabry Perot optical modes, $a$ and $c$, are coupled to the same mechanical mode, $b$. (c) Frequency-domain illustration of the control and probe lasers used throughout this work, and their respective optical cavity modes. (d) The double-$\Lambda$ level scheme used in this work, where the labels $n_\text{a}, n_\text{b}, n_\text{c}$ represent the energy levels of modes $a$, $b$, and $c$, respectively. The thick lines represent photon-phonon exchange mediated by the control field, and the thin lines represent the probe fields with phase difference $\phi=\phi_\text{a}-\phi_\text{c}$. (e) DOMIT can be described by three coupled harmonic oscillators. The optical modes may be driven in-phase-, or out-of-phase, exciting either the mechanically bright or the mechanically dark mode.}
\label{fig:1}
\end{figure}

Here we utilize a cavity optomechanical device with two optical modes coherently coupled to a single mechanical resonance to show that double optomechanically induced transparency (DOMIT), in which two optical modes coherently couple to the same mechanical mode, enables interference between photons separated by over 5 THz in frequency. Exploiting this effect, we demonstrate a phase-sensitive two-colour optical XOR gate \cite{ref:zhang2005aox}. This multi-colour interference is mediated by the mechanical bright mode and the mechanical dark mode \cite{ref:dong2012ODM, ref:wang2012uif}, which are directly excited for the first time here. This builds on previous studies using multimode optomechanical systems, which achieved partial dark mode excitation. In these works, which were focused on wavelength conversion and thus involved a single probe, finite perfect dark mode excitation can only be realized in the limit of infinite optomechanical cooperativity, $C$ \cite{ref:dong2012ODM}. The optical XOR gate demonstrated here will be useful for multicolour classical and quantum optical information processing, e.g., for demonstrating quantum interference \cite{ref:kobayashi2016frequency, ref:li2019tunable} between frequency binned qubits \cite{ref:clemmen2016ramsey, ref:kues2017chip, ref:lu2018quantum}, as well as optical sensing and metrology \cite{ref:Qu2019CBD} using interference between widely separated wavelengths.

\section*{Results}


\noindent\textbf{Double optomechanically induced transparency (DOMIT).} Photon-photon interference and mechanical dark mode excitation demonstrated here uses a diamond microdisk cavity optomechanical system (Fig.\ \ref{fig:1}(a)) whose essential elements can be analyzed as a generic Fabry Perot cavity shown in Fig.\ \ref{fig:1}(b). The cavity supports two optical modes with widely separated frequencies $\omega_\text{a}$ and $\omega_\text{c}$, each coupled via radiation pressure to a common mechanical resonator with frequency $\omega_\text{b}$ established in the generic system by one of the cavity mirrors. Each optical mode is excited with a weak probe field and a strong control laser that is red-detuned from its cavity mode by the mechanical frequency, $\Delta_\text{a} = \Delta_\text{c}=-\omega_\text{b}$, represented graphically in Fig.\ \ref{fig:1}(c). If the mechanical frequency exceeds the dissipation rate of each optical cavity mode, the system is in the resolved sideband regime, and the interaction Hamiltonian simplifies to \cite{ref:aspelmeyer2014co}
\begin{equation}\label{eq:Hamiltonian}
\hat{H}_\text{int} = -\hbar\left[\left (G_\text{a}\hat{a}^\dagger+G_\text{c}\hat{c}^\dagger \right)\hat{b} + \left(G_\text{a}^{\ast}\hat{a}+ G_\text{c}^{*}\hat{c} \right)\hat{b}^\dagger\right],
\end{equation}
\noindent where $\hat{a}\, (\hat{a}^\dagger)$ and $\hat{c}\,(\hat{c}^\dagger)$ are the creation (annihilation) operators of the optical probe field photons, $\hat{b}\, (\hat{b}^\dagger)$ is the creation (annihilation) operator of the mechanical resonator phonons, and we employ the rotating wave approximation. Here $G_\text{a}=g_\text{a}\alpha_\text{a}$ and $G_\text{c}=g_\text{c}\alpha_\text{c}$ are the control- field assisted optomechanical coupling rates for modes $a$ and $c$, set by the single photon-phonon coupling rates $g_\text{a,c}$, and the control field amplitudes $|\alpha_\text{a,c}|^2 = n_\text{a,c}$. The specific cavity geometry determines $g_\text{a,c}$, which generally increases as the effective cavity length decreases, whereas the control field amplitudes are set by their intracavity photon numbers $n_\text{a,c}$, which are typically limited by nonlinear optical effects in the cavity. The system can be described by a double-$\Lambda$ energy diagram, as illustrated in Fig.\ \ref{fig:1}(d), which forms a closed loop under excitation from the two sets of control and probe fields.

In this DOMIT configuration, each of $\hat{a}$, $\hat{b}$, and $\hat{c}$ oscillate in the rotating frame at the same frequency. This allows relative phases between the various fields to be defined, despite their typically vast frequency differences. The key property is that the phases of each of the four fields forming the double-$\Lambda$ loop affects the optical response \cite{ref:buckle1986ai, ref:kosachiov1992cpm}. This is in contrast to a single-$\Lambda$ system, whose optical response depends only on the intensity of the control field. This behaviour is clearly elucidated by studying symmetric and antisymmetric combinations of the cavity's optical modes. These modes are referred to as the `mechanically dark' $\left(\hat{\zeta}_\text{dk} = (G_\text{c} \hat{a} - G_\text{a} \hat{c})/(\1i\overline{G})\right)$ and `mechanically bright' $\left(\hat{\zeta}_\text{br} = (G^{*}_\text{a} \hat{a} + G^{*}_\text{c} \hat{c} )/(\overline{G})\right)$ modes, as the dark mode can be entirely decoupled from the mechanical resonator while the bright mode can be maximally coupled; here $\overline{G} = \sqrt{|G_\text{a}|^2+|G_\text{c}|^2}$.

These three modes are analogous to the modes of three coupled pendula, as shown in Fig.\ \ref{fig:1}(e), in which the outer two pendula move in opposite directions while the central pendulum is stationary (dark mode); alternatively the three pendula move in the same direction (bright mode). This basis, with analogies in atomic physics \cite{ref:korsunsky1999phase, ref:Kim2015ALS}, elegantly reveals the importance of optical phase to the system's behaviour. The classical amplitudes of the bright and dark modes when both probe fields are resonant with their respective cavity modes are (see Supplementary Note 4):
\begin{align}\label{eq:modeSelection1}
\zeta_\text{dk} &= \frac{2\sqrt{\kappa_\text{ex}}s_\text{in}}{\kappa}\sin\left(\phi/2\right), \\ \label{eq:modeSelection2}
\zeta_\text{br} &= \frac{2\sqrt{\kappa_\text{ex}}s_\text{in}}{\kappa \left(1+\overline{C}\right)}\cos\left(\phi/2\right),
\end{align}
where $\kappa_\text{ex}$ and $\kappa$ are the external coupling and total loss rates of the cavity modes respectively, $s_\text{in}$ is the input amplitude of the two probe lasers, $\overline{C}=4\overline{G}^2/\kappa\gamma_\text{b}$ is the two-mode optomechanical cooperativity, and $\gamma_\text{b}$ is the damping rate of the mechanical resonance. For simplicity, we assume that $\kappa$, $\kappa_\text{ex}$ are the same for each optical mode, and set $s_\text{in}$ to be equal to reflect the fact that we balanced the input probe field amplitudes in the experiment. In our experimental setup described below, $\phi_\text{a}$ is the phase difference between control and probe fields input to mode $a$ and $\phi_\text{c}$ is the phase difference between control and probe fields input to mode $c$. The total phase difference is then $\phi=\phi_\text{a}-\phi_\text{c}$. As each of the probe fields are derived from their respective control field via electro-optic modulation in our experiment, changes to the control field phases do not affect the system response because they will also shift the probe phase by the same amount. Note that in principle the system could be operated with any of the four control and probe phases used to determine the nature of the interference.

Equations (\ref{eq:modeSelection1}-\ref{eq:modeSelection2}) show that adjusting $\phi$ to $\pm\pi$ e.g., by delaying one of the probes, allows complete selective excitation of $\zeta_\text{dk}$ without requiring $\overline{C}\rightarrow \infty$. When all DOMIT processes are resonant, this delay corresponds to a half period of the mechanical resonator. Equations (\ref{eq:modeSelection1}-\ref{eq:modeSelection2}) also show that full optical power transfer to the dark mode is possible, while the bright mode has a reduced maximum amplitude due to coherent optomechanical transfer of energy to the mechanical resonator. Optically manipulating the system in this basis is central to optomechanical wavelength conversion free from mechanical thermal decoherence effects \cite{ref:wang2012uif, ref:tian2012asc} when the optomechanical coupling exceeds the thermal decoherence rate \cite{ref:palomaki2013cst, ref:hill2012cow, ref:dong2012ODM}. However, none of these previous single OMIT studies have completely isolated or selectively populated the mechanically dark mode. Conceptually related studies demonstrating optomechanical control of interference between two mechanical modes \cite{ref:Kuzyk2017MultiInt} have not yet been used to interfere different colours of light.

\begin{figure*}
\centering
\includegraphics[width=\linewidth]{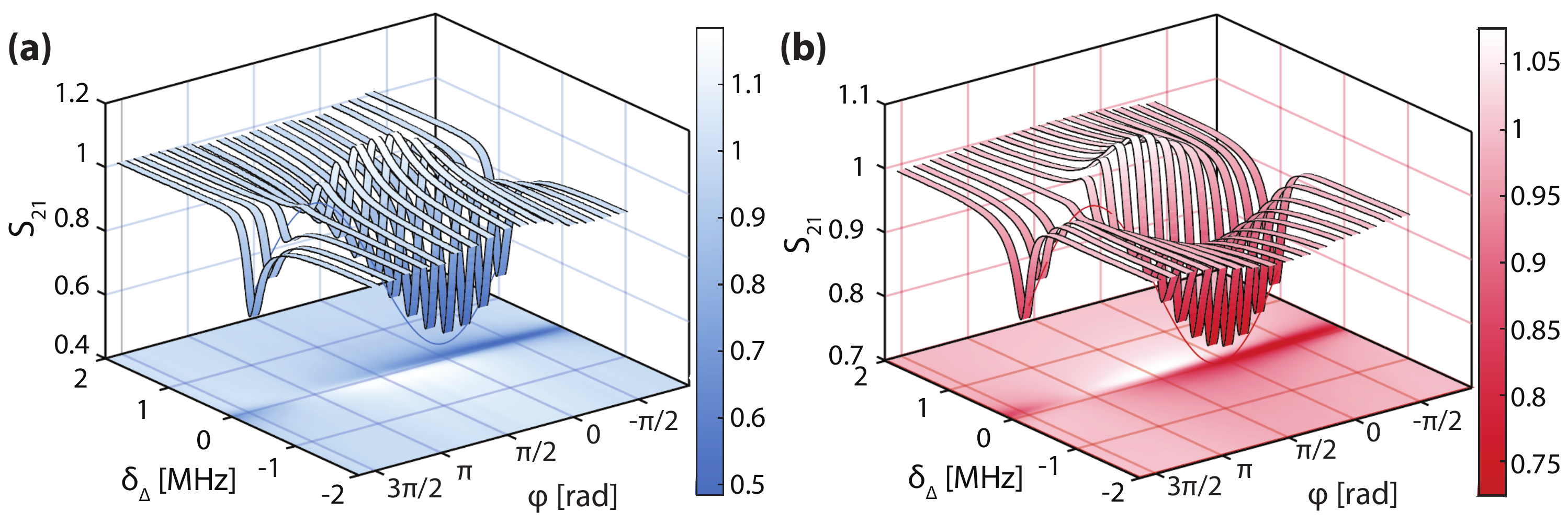}
\caption{Measurement of double optomechanically induced transparency. (a,b) Cavity optical response to the probe signal for modes ${a}$ and ${c}$ respectively, as a function of cavity-probe detuning $\delta_\Delta$ and the probe phase difference $\phi$, as recorded by the vector network analyzer from the photodetected heterodyne signal. The control fields are at fixed detuning $\Delta_\text{a} = \Delta_\text{c} = -\omega_\text{b}$.}
\label{fig:2}
\end{figure*}

To demonstrate DOMIT, we evanescently coupled control and probe fields via an optical fiber taper waveguide into modes of a diamond microdisk device similar to that in Fig.\,\ref{fig:1}(a) and previous studies \cite{ref:mitchell2016scd, ref:lake2018oit}. As shown in Supplementary Fig.\,1, our modes have resonant wavelengths $\lambda_\text{a}\sim 1520$ nm and $\lambda_\text{c} \sim 1560$ nm and sufficiently low optical loss ($\kappa_\text{a}/2\pi \sim 0.87$ GHz, $\kappa_\text{c}/2\pi \sim 1.20$ GHz) to allow resolved sideband optomechanical coupling to the microdisk's $\omega_\text{b}/2\pi = 2.1$ GHz fundamental mechanical radial breathing mode (dissipation rate $\gamma_\text{b}/2\pi = 0.285 $ MHz). Typical optical and mechanical mode spectroscopy measurements are described in Supplementary Note 1. The per-photon optomechanical coupling rates, $g_{\text{a}} = g_{\text{c}}\sim 2\pi \times 25$ kHz, allow optomechanical cooperativity $> 1$ and observation of OMIT when approximately $n_\text{a,b} > 5 \times 10^5$ control photons are coupled into either of the cavity modes \cite{ref:lake2018oit}. Achieving this large photon number is possible in our microdisks due to diamond's low nonlinear absorption and excellent thermal properties. In all of our measurements presented below, the input control fields are detuned from their respective cavity modes such that OMIT conditions $\Delta_\text{a} = \Delta_\text{c} = -\omega_\text{b}$ for each mode are satisfied.  Note that the microdisks support regularly spaced modes spanning the IR and visible spectrum \cite{ref:mitchell2016scd}; our modes are chosen due to the compatibility of their wavelengths with telecommunications equipment needed for the measurements described below.

To excite $\zeta_\text{dk}$ and $\zeta_\text{br}$, OMIT spectra \cite{ref:lake2018oit} for modes $a$ and $c$ were recorded for varying relative phase $\phi$ between the two probe fields, as shown in Figs.\ \ref{fig:2}(a,b). In these measurements, $\phi$ is controlled by adjusting a delay between two radio frequency (RF) signals split from the same signal generator, which are then used to create the probe fields through optical modulation of two independently running control lasers as shown in Supplementary Fig.\,2 (see Methods and Supplementary Note 1 for details). Our setup is robust to phase drifts of either control laser, as discussed above. Each pair of probe and control fields is isolated by optical filtering and then detected on a high-speed photodetector as a function of varying probe-cavity detuning $\delta_\Delta$. Each probe amplitude is then measured by downmixing the heterodyne signal that it creates through interference with its corresponding control field using a vector network analyzer. At $\phi=0$, the deep OMIT window present in each probe output spectrum when $\delta_\Delta = 0$ indicates excitation of $\zeta_\text{br}$, whereas, when $\phi=\pi$, the response of the bare cavity, which is broad compared to the OMIT window ($\kappa \gg \gamma_\text{b}(1 + C)$), is restored for all probe detunings, indicating excitation of $\zeta_\text{dk}$. The depth of the OMIT window as a function of the phase delay is fit using solutions to Eqns.\,(\ref{eq:modeSelection1}-\ref{eq:modeSelection2}) in the $a$ and $c$ basis (see Supplementary Note 4), as is shown for $\delta_\Delta^\text{a,c} = 0$ in Figs.\,\ref{fig:2}(a,b), where it is seen to agree well with theory. The OMIT line-shapes of each mode deviate from a Lorentzian when studied individually \cite{ref:weis2010oit, ref:safavi2011eit} due to the use of an amplitude modulator with non-zero chirp (see Supplementary Note 2 and Supplementary Fig.\,3) in generating one of the probe lasers, and imperfect detuning of the control fields, as analyzed in previous work \cite{ref:mitchell2019oaw}.  Differences in the Fano phase of each OMIT line-shape then results in constructive or destructive interference features for non-zero $\delta_\Delta$ visible in Figs.\ \ref{fig:2}(a,b).


\noindent \textbf{Optomechanical bright and dark mode coupling.} When OMIT resonance conditions for both modes are satisfied, i.e., $\Delta_\text{a} = \Delta_\text{c} = -\omega_\text{b}$ and $\delta_\Delta^\text{a} = \delta_\Delta^\text{c} = 0$, the $\zeta_\text{br}$ and $\zeta_\text{dk}$ states are decoupled from each other. However, as shown on Supplementary Fig.\,5, by incrementing cavity-probe detuning of one mode by $+\delta_\Delta$, and decrementing cavity-probe detuning of the other mode by $-\delta_\Delta$, we induce a bright-dark state coupling. This can also be accomplished by shifting $\Delta_\text{a}$ and $\Delta_\text{c}$ in the same manner. Coupling between bright and dark states manifests as a temporal oscillation in the intensity of the probe fields transmitted through the cavity, allowing differences in their interaction with the dissipative mechanical resonator to be observed directly (see Supplementary Note 3).

\begin{figure*}
\centering
\includegraphics[width=\linewidth]{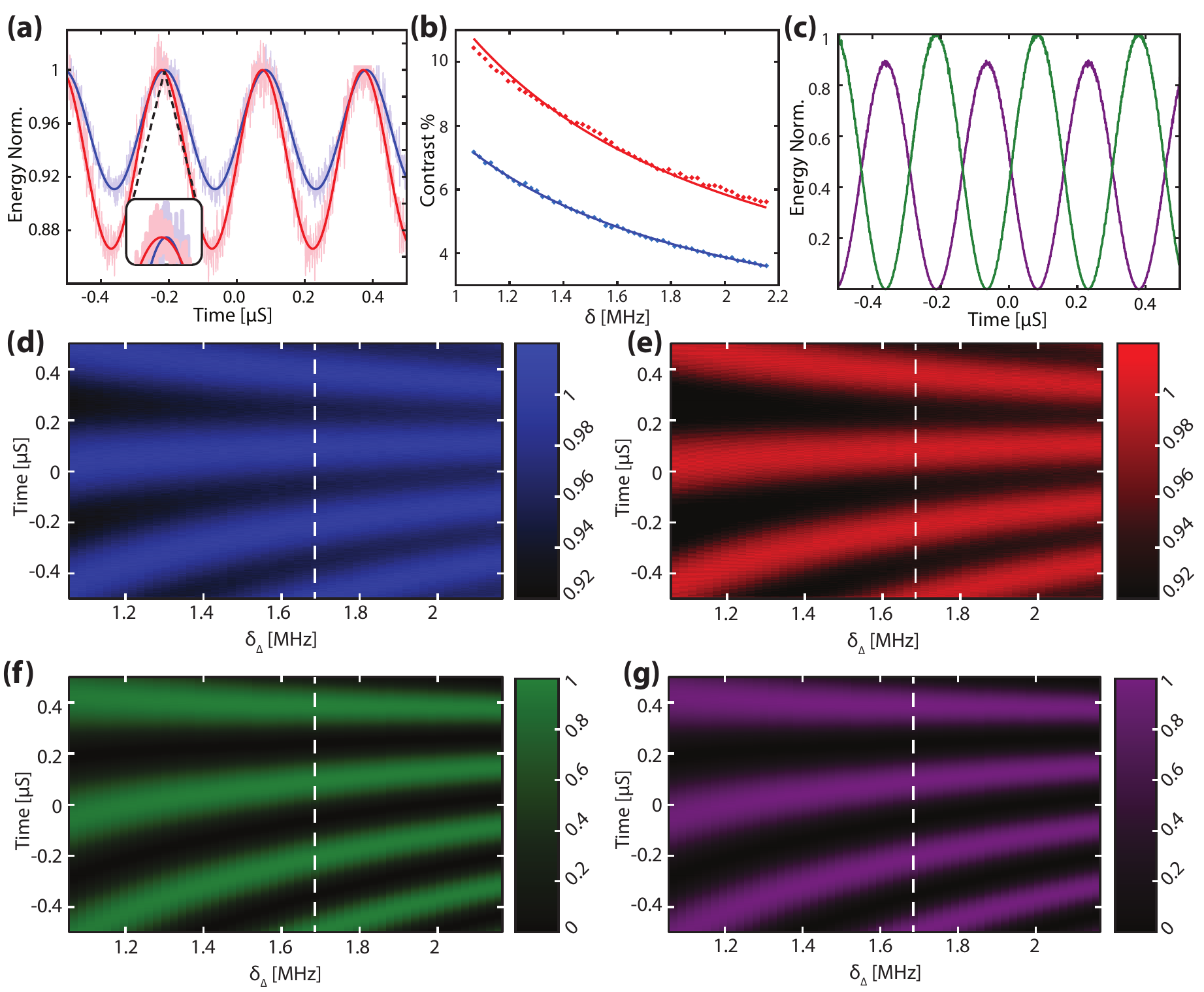}
\caption{Measurement of coupling between the mechanically dark and bright modes. (a) Example of the oscillations in stored energy in mode $a$ (blue trace) and mode $c$ (red trace) for $2\delta_\Delta = 3.37$ MHz, measured by downmixing the heterodyne signal at each colour in the time domain to isolate each probe amplitude. Inset: highlight of the slight time delay between the output of modes $a$ and $c$ caused by dispersion in the setup. (b) Amplitude of the oscillations for mode $a$ (blue) and mode $c$ (red) as a function of $\delta_\Delta$, and corresponding predictions from the model given in the text. (c) Normalized bright (purple) and dark (green) state energy for $2\delta_\Delta = 3.37$ MHz, as inferred from the optical output of modes $a$ and $c$ shown in (a). (d-g) Oscillations as a function of time and $\delta_\Delta$ of outputs from modes $a$, $c$, $\zeta_\text{br}$, and $\zeta_\text{dk}$ respectively.}
\label{fig:3}
\end{figure*}

Our measurement of both probe colours is plotted in Fig.\ \ref{fig:3}(a), for the case that $2\delta_\Delta=3.37~\text{MHz}$ and $\phi=0$, after digitally downmixing the total (probe and control) photodetected signal from each colour recorded on a high-speed oscilloscope to remove fast oscillations near $\omega_\text{b}$ from beating between probe and control fields. Each downmixed signal is proportional to the amplitude, or equivalently, the square root of energy of the intracavity field at its respective probe frequency. As the modulation depth is bounded above by the dark-state transmission (bare cavity response), and bounded below by the bright-state transmission (OMIT window depth), the oscillation amplitude follows the $\delta_\Delta$ dependence of the OMIT features. This dependence is confirmed by measuring the dependence of oscillation amplitude on increasing $\delta_\Delta$, as shown in Fig.\ \ref{fig:3}(b), which matches well with theoretical predictions (see Supplementary Notes 5 and 6 for details), giving estimates of $\overline{C} = 3.6$ and $\overline{C} = 4.2$ from fits to the $\omega_\text{a}$ and $\omega_\text{c}$ photodetected signals, respectively. This difference in the cooperativities is attributed to non-ideal modulation when creating each probe and imbalance of the cooperativities of each mode.

We can also reconstruct the output of ${\zeta}_\text{dk}$ and ${\zeta}_\text{br}$ by inferring $a$ and $c$ from the measured phase and amplitude of the output probe fields in Fig.\ \ref{fig:3}(a). This reconstruction requires the additional step of accounting for a slight phase shift due to differing optical path lengths at $\omega_\text{a}$ and $\omega_\text{c}$ caused by dispersion in the experimental setup, whose effect can be seen in the inset of Fig.\ \ref{fig:3}(a). This deleterious phase shift was corrected in post-processing while determining ${\zeta}_\text{br}$ and ${\zeta}_\text{dk}$. An example of the reconstructed bright and dark mode intensity is shown in Fig.~\ref{fig:3}(c), where flopping between the bright and dark state is evident. Notably, a difference in maximum intensity of $\zeta_\text{dk}$ and $\zeta_\text{br}$ is evident from their differing peak values.
This difference can be related to the energy dissipated by the bright state due to its interaction with the mechanical resonance, and is found from Eqns.\ (\ref{eq:modeSelection1}-\ref{eq:modeSelection2}) generalized for non-zero $\delta_\Delta$ (see Supplementary Notes 5 and 6) to scale as $(1 + \overline{C}/(1 + 4(\delta_\Delta/\gamma_\text{b})^2)^{-2}$. Additional measurements of the intensity of $a$, $c$, $\zeta_\text{br}$, and $\zeta_{dk}$ as a function of $\delta_\Delta$ and time are plotted in Figs.\ \ref{fig:3}(d--g), which clearly show how the oscillation period decreases with increasing $\delta_\Delta$, as expected theoretically. The effect of mismatch in operating parameters between mode $a$ and $c$ is calculated in Supplementary Note 5 and Supplementary Fig.\,4, where the main effect of mismatch in optomechanical coupling or probe amplitudes is to decrease the contrast of the interference, whereas mismatch in frequency or optical decay rates leads to dissipative and dispersive coupling between the bright and dark states.

\noindent \textbf{Two-colour switching.} The phase-dependent response of this multicolour DOMIT system, together with our ability to selectively excite $\zeta_\text{dk}$ or $\zeta_\text{br}$ can be harnessed to create a phase-dependent all-optical switch. In this device the output intensity of one probe is dependent on the phase of the other probe, and follows the truth table of an XOR gate with probe field phases of 0 and $\pi$ mapping onto Boolean values 0 and 1. The maximum contrast achievable is determined by the maximum OMIT dip depth, and is given by $\overline{C}^2/(1+\overline{C})^2$. This indicates that, in principle, the contrast can be made to approach unity for systems with large cooperativity. The main advantage of the switching scheme demonstrated here compared to Ref.\cite{ref:Kuzyk2017MultiInt} is that the switch inputs and outputs are all-optical, and the scale of frequency differences in our systems is large, i.e. the optical modes in our system are separated by 5 THz, compared with only 180 kHz of the nearly degenerate mechanical modes in Ref.\cite{ref:Kuzyk2017MultiInt}.

\begin{figure}[ht]
\centering
\includegraphics[width=\linewidth]{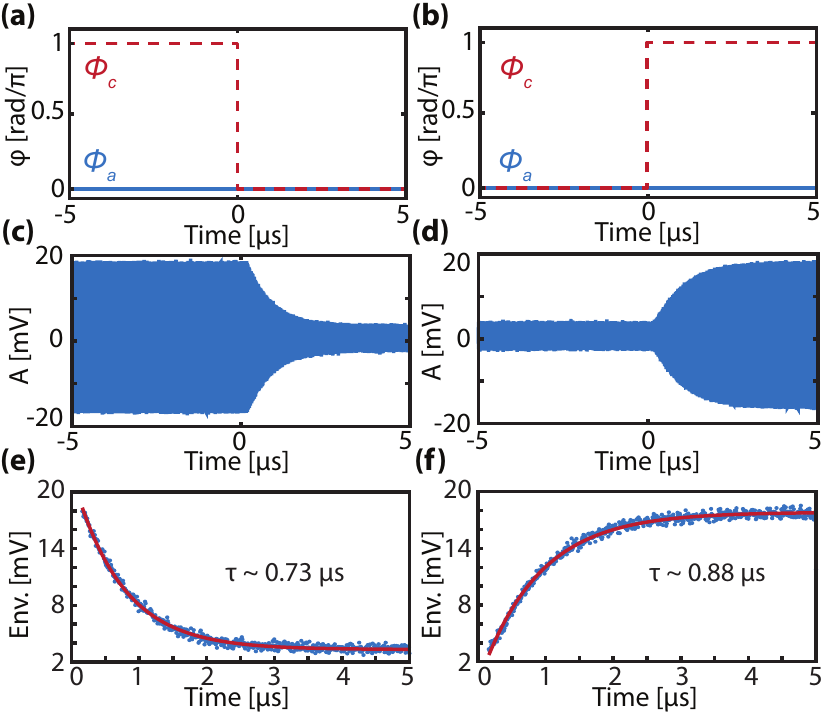}
\caption{Demonstration of a phase-dependent all-optical switch. (a,b) Phases of the input probe fields vs.\ time which cause probe $c$ to switch the response of probe $a$ off and on, respectively. (c,d) Photodetected cavity output of probe $a$ when the input phase of probe $c$ is switched temporally as shown in (a) and (b). The rapid oscillations are caused by beating between the control and probe, and the envelope reveals the action of the switch. (e,f) Envelope extracted from the peak values of the data in (c,d) and fit with exponentially falling and rising functions, respectively.}
\label{fig:4}
\end{figure}

This switching action can be inferred from Figs.\ \ref{fig:2}(a,b), and directly observed in the time domain by varying the phase $\phi_\text{c}$ of the mode $c$ probe following a temporal step function, while maintaining constant phase $\phi_\text{a}$ of the mode $a$ probe, as sketched in Fig.\ \ref{fig:4}(a) for switching off and Fig.\ \ref{fig:4}(b) for switching on. Experimental time domain data showing the resulting change in probe $a$ transmission is shown in Figs.\ \ref{fig:4}(c,d). As mentioned above, the relative phase $\phi = \phi_\text{a} - \phi_\text{c}$ is controlled by introducing an electrical phase delay in the RF signal driving the electro-optic modulator responsible for creating the mode $c$ probe (see Methods and Supplementary Note 1 for details). The rapid oscillations are due to beating between the probe and control laser, whereas the oscillation envelope is proportional to the amplitude of the transmitted probe. From this upper envelope, shown in
Figs.\ \ref{fig:4}(e,f), we can measure the switch response speeds: fits with an exponential function yield fall and rise times of $0.73 \ \mu\text{s}$ and $0.88 \ \mu\text{s}$ respectively. Assuming that $\kappa_\text{a,c}\gg\gamma_\text{b}$, the switching speed can be shown to be $\tau=1/\gamma_\text{b}(1+\overline{C})$ (see Supplementary Note 7), which approaches zero for sufficiently large $\overline{C}$, enhancing the switching speed beyond $1/\gamma_\text{b}$. This switching speed implies $\overline{C} = 2.96$ and $\overline{C} =3.89$ respectively, for the fall and rise times illustrated in Figs.\ \ref{fig:4}(c,d), consistent with expectations from the control field amplitudes. The observed differences in the fall and rise times are presumed to be due to differences in intracavity control field amplitudes during the respective measurements. Larger contrast, which would be beneficial for switching applications, could be realized by further increasing $C$. Single photon operation would require removal of thermal phonons, either through cryogenic cooling or higher cooperativity combined with feedforward noise suppression \cite{ref:higginbotham2018heo}.

\section*{Discussion}

\noindent In summary, we have demonstrated coherent interference between spectrally-separated optical modes mediated by optomechanical coupling. By adjusting the phase between different colour probe fields entering the cavity, we selectively excite either a mechanically bright or a mechanically dark mode, and demonstrate controllable coupling between the two modes. Notably, we exploit the difference between bright- and dark-state transmission to demonstrate a two-colour, all-optical switch, where the control and target are at different wavelengths.

Our system has great potential for applications to quantum information processing where interference between frequency binned qubits is desirable \cite{ref:lu2018quantum}, such as frequency-domain Hong-Ou-Mandel interference \cite{ref:kobayashi2016frequency, ref:li2019tunable}, chromatic and time-domain interferometry \cite{ref:Qu2019CBD, ref:reddy2018high, ref:clemmen2016ramsey}, and microwave-to-optical conversion via the optomechanical dark mode \cite{ref:balram2016ccr}. Furthermore, as detailed in Supplementary Note 8, our technique could be extended to many optical modes whose operating wavelengths are only limited by the transparency of diamond and the existence of high-$Q$ modes; thus, our technique could lead to many-colour interference processes. Finally, we note that this interference is quite general and could be utilized for non-optical inputs such as magnetic or electric fields, provided they couple to the mechanical degree of freedom \cite{ref:wu2016not}. Future experiments operating in the quantum domain will benefit from cryogenic pre-cooling of the device to its mechanical quantum ground state, which is achievable at dilution fridge temperatures for the $> 2~\text{GHz}$ frequency mechanical mode measured in this article.

\section*{Methods}

\noindent\textbf{Device.} The device used for all experiments in this work is a $\sim5 \,\mu\text{m}$ microdisk resonator fabricated from single crystal diamond using the process outlined in \cite{ref:khanaliloo2015hqv,ref:mitchell2018rqo}. A full schematic of the experimental apparatus used in the experiments is given in Supplementary Note 1. In all of the measurements light was coupled evanescently into and out of the microdisk using a dimpled optical fiber taper positioned using motorized stages, as discussed in \cite{ref:mitchell2016scd, ref:lake2018oit}.

\noindent\textbf{Measurement of DOMIT.} To measure the OMIT spectra of each microdisk mode, sidebands were created on each respective control laser for use as probe fields, using either phase, $\phi(t)$, or amplitude, $A(t)$, electro-optic modulators. For the data in Fig.\,2, the electrical RF drive for each modulator was derived from the same vector network analyzer, with one path undergoing a controllable phase shift relative to the other. This controllable phase shift was achieved by placing an electronic phase shifter before one of the electro--optic modulators. Because the phase shifter transmission varied as a function of phase, a variable electrical attenuator was calibrated and used to maintain balance between the probe laser powers at every step in phase, by controlling the RF modulation amplitude. An optical variable attenuator was used on the 1560 nm laser arm to attempt to balance the input power of each laser before they were combined via a 50/50 waveguide coupler, and amplified with an erbium doped amplifier before being routed to the sample chamber (a nitrogen purged enclosure) and device. The other output of the 50/50 coupler was used to perform slow laser wavelength locking via a photodetector and optical spectrum analyzer connected to the control computer. The signal exiting the sample chamber was then divided on a 90/10 waveguide coupler. The 10\% port was routed to a low speed photodetector for use in measuring the cavity transmission during the initial setup, and the 90\% port was sent to the wavelength division multiplexer (WDM). By connecting the WDM to a high-speed photodetector, the output of either mode $a$ or $c$ could be selected.

\noindent\textbf{Measurement of bright--dark mode coupling and switching.} For the bright--dark mode coupling experiment (Fig.\,3) and the time domain switch (Fig.\,4), a two--channel arbitrary waveform generator (AWG) was used as the RF source, with one channel assigned to each modulator. Acquisition was performed using a digital spectrum analyzer (DSA), which was triggered by the AWG. To isolate the beat note between the probe field of one mode and the converted probe from the other mode the signal acquired on the DSA was digitally downmixed post--acquisition.

\section*{Data availability}
\noindent The datasets generated during and/or analyzed during the current study are available from the corresponding author on reasonable request

\section*{Acknowledgments}
\noindent This work was supported by the NRC Nanotechnology Research Centre Nano Initiative, Canadian Foundation for Innovation, Alberta Innovates, and the Natural Sciences and Engineering Research Council of Canada.

\section*{Author contributions}

\noindent D.P.L. and M.M. fabricated the device, performed the experiments, and analyzed the data. D.P.L. developed the theory, and D.P.L., M.M., B.C.S., and P.E.B. contributed to writing the manuscript.

\section*{Competing interests}

\noindent The authors declare no competing interests.

\section*{References}

\clearpage

\newpage

\setcounter{equation}{0}
\setcounter{figure}{0}
\setcounter{section}{0}
\setcounter{subsection}{0}
\setcounter{table}{0}
\setcounter{page}{1}
\makeatletter
\renewcommand{\theequation}{S\arabic{equation}}
\renewcommand{\thetable}{S\arabic{table}}
\renewcommand{\thefigure}{S\arabic{figure}}
\renewcommand{ \citenumfont}[1]{S#1}
\renewcommand{\bibnumfmt}[1]{[S#1]}

\onecolumngrid

\section{Supplementary Information}

\section*{Supplementary Note 1 - Experiment setup and calibration}

The optomechanical cavity utilized in this work is a single--crystal diamond (SCD) microdisk, fabricated according to the process outlined in Refs.\ \cite{ref:supp_khanaliloo2015hqv,ref:supp_mitchell2018rqo}, an example of which is shown in Supplementary Figure \ref{fig:s1}(a). An advantage of microdisk cavities is that they support multiple optical whispering gallery modes across their transparency window, all of which exhibit dispersive optomechanical coupling to the fundamental radial breathing mode (RBM) of the microdisk \cite{ref:supp_lake2018oit}, as illustrated schematically in Supplementary Figure \ref{fig:s1}(b). Diamond's large electronic bandgap, Young's modulus, and best-in-class thermal conductivity make it an ideal material for use in cavity optomechanics as it can support large intracavity photon number $N$, and high optical and mechanical quality factors. Additionally, colour center qubits present in diamond, such as silicon and nitrogen vacancies, make it a promising platform for realizing hybrid quantum systems \cite{ref:supp_aharonovich2011dp}.

\begin{figure}[ht]
\centering
\includegraphics[width=\linewidth]{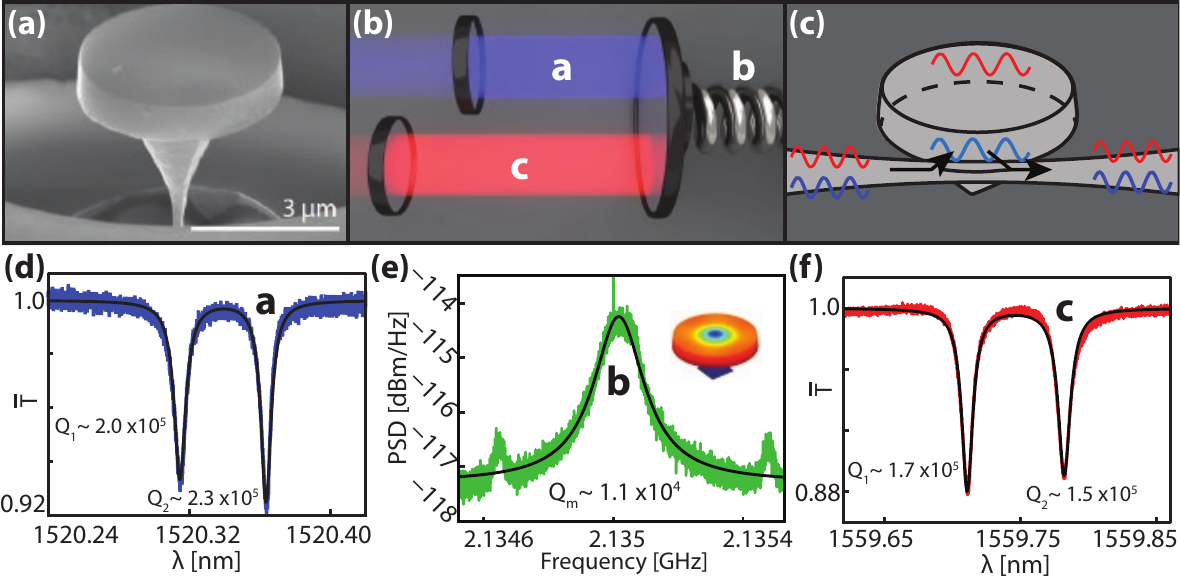}
\caption{Characterization of optical and mechanical modes of the diamond microdisk utilized in this work. (a) Scanning electron micrograph of a diamond microdisk similar to the one used in this work. (b) Cartoon of a canonical multimode optomechanical system. Here mode $b$, represented as a spring, is dispersively coupled to both optical modes $a$ and $c$. (c) Cartoon of fiber taper--microdisk coupling. (d) Normalized fiber taper transmission scan of mode $a$ used in this experiment, with fit. (e) Power spectral density of the fiber taper transmission when the input laser is tuned near a cavity mode, revealing fluctuations from thermomechanical motion of the cavity's mechanical radial breathing mode, $b$. A COMSOL simulated displacement field profile of the radial breathing mode is shown in the inset. (f) Normalized fiber taper transmission scan of optical mode $c$, with fit.}
\label{fig:s1}
\end{figure}

Light from two tunable diode lasers was coupled into and out of the microdisk using a dimpled optical fiber taper positioned adjacent to the microdisk as illustrated in the cartoon in Supplementary Figure \ref{fig:s1}(c). The spatial overlap of the evanescent field of the fiber and the optical modes of the microdisk permit efficient coupling, allowing measurement of cavity modes in transmission and reflection. Two telecommunications wavelength modes at $\lambda_\text{a} = 2 \pi c / \omega_\text{a} = 1520$ nm and $\lambda_\text{c} = 2 \pi c / \omega_\text{c} = 1560$ nm were selected for this work, as they were in the operating range of the available lasers (Newport TLB-6700) and optical amplifier (Pritel EDFA). However, this could be extended to visible wavelengths, where these devices have demonstrated high quality optical modes \cite{ref:supp_mitchell2016scd}. The optical modes are each dispersively coupled to the microdisk's fundamental mechanical radial breathing mode (RBM) whose frequency is $\omega_\text{b}/2\pi = 2.1$ GHz, with vacuum optomechanical coupling rates, $g_{0,\text{a}},g_{0,\text{c}}\sim 2\pi\times25$ kHz. Measurements of the fiber taper optical transmission spectrum for wavelengths scanned across modes $a$ and $c$ are shown in Supplementary Figure \ref{fig:s1}(d,f), and the power spectral density of the fluctuations imparted on photodetected output due to thermally driven mechanical motion of the RBM when the input laser is tuned close to resonance with an optical mode \cite{ref:supp_kippenberg2005arp, ref:supp_carmon2005temporal, ref:supp_rokhsari2005radiation} is shown in Supplementary Figure \ref{fig:s1}(e). Note that both optical modes are standing wave doublets due to surface roughness induced coupling between the clockwise and counterclockwise propagating whispering gallery modes of the microdisks \cite{ref:supp_kippenberg2002mct}. For all of the measurements presented here the long wavelength doublet mode was used, and the lower wavelength doublet mode was assumed to not influence the observed phenomena. This device operates in the sideband resolved regime, $\omega_\text{b} \gg \kappa_\text{a},\kappa_\text{c}$, where $\kappa_\text{a}/2\pi \sim 0.87$ GHz, and $\kappa_\text{c}/2\pi \sim 1.20$ GHz are the energy decay rates of the optical modes. A previous study \cite{ref:supp_mitchell2018rqo} found that the optical quality factor, $Q = \omega/\kappa$, of this device is likely still limited by surface roughness induced by the fabrication procedure.


\begin{figure}[ht]
\centering
\includegraphics[width=\linewidth]{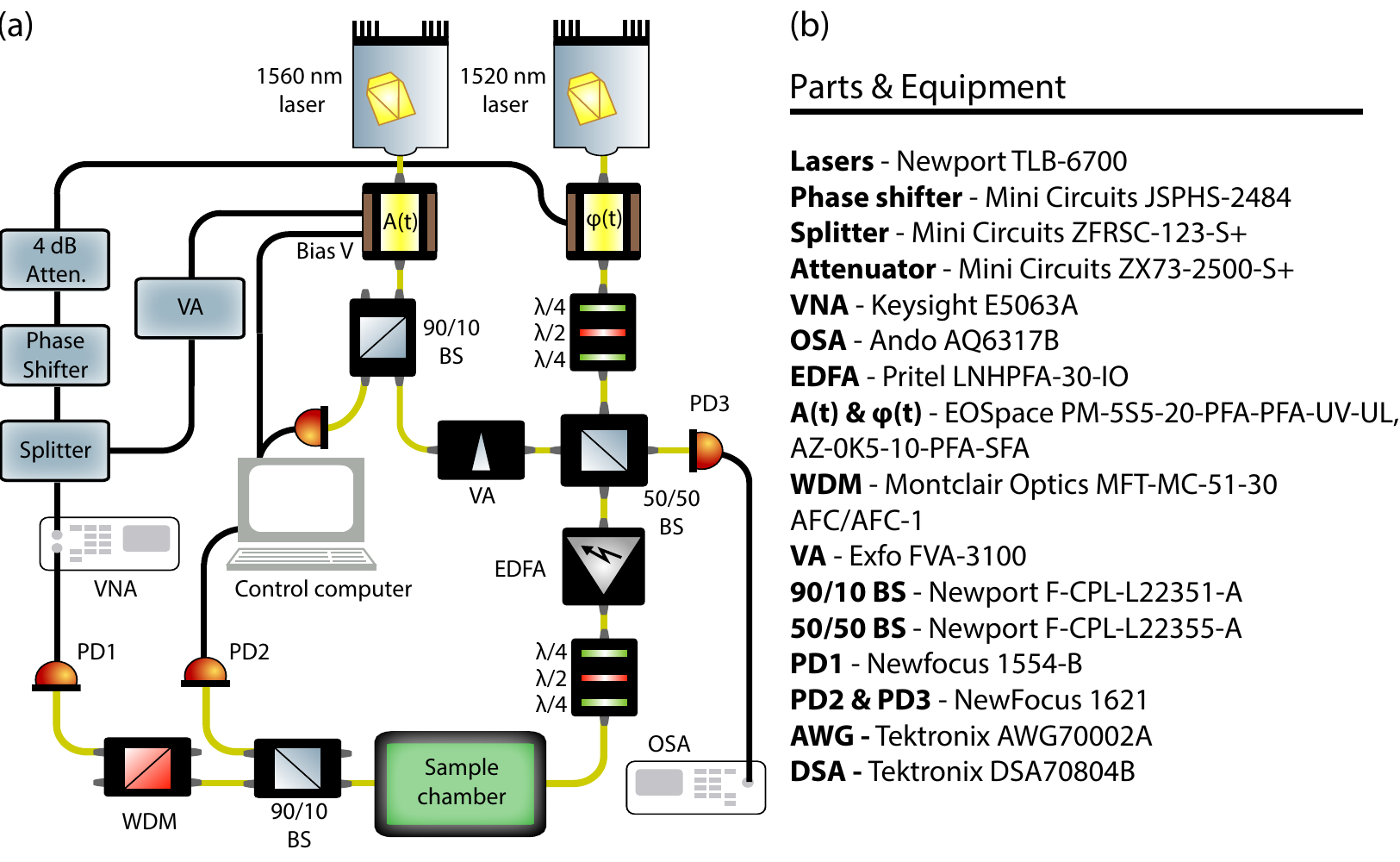}
\caption{Apparatus used in this work, with corresponding equipment list. (a) Apparatus used in the experiments. The Vector Analyzer (VNA) acted as the RF source for modulators on each path. The resulting signal was spectrally filtered to isolate for the mode of interest. Key: OSA (Optical spectrum analyzer), BS (Beam splitter), VA (Variable attenuator), BS (Beam splitter), EDFA (Erbium doped fiber amplifier), WDM (Wavelength division multiplexer), PD (Photodetector). (b) Parts and equipment used in the experiment.}
\label{fig:s2}
\end{figure}

Supplementary Figure \ref{fig:s2} shows a schematic representation of the experimental setup used for the above measurements and those shown in the main text. Additional information can be found in the methods section of the main text.

\section*{Supplementary Note 2 - Electro--optic modulation and probe measurement model}

\begin{figure}[ht]
\centering
\includegraphics[width=\linewidth]{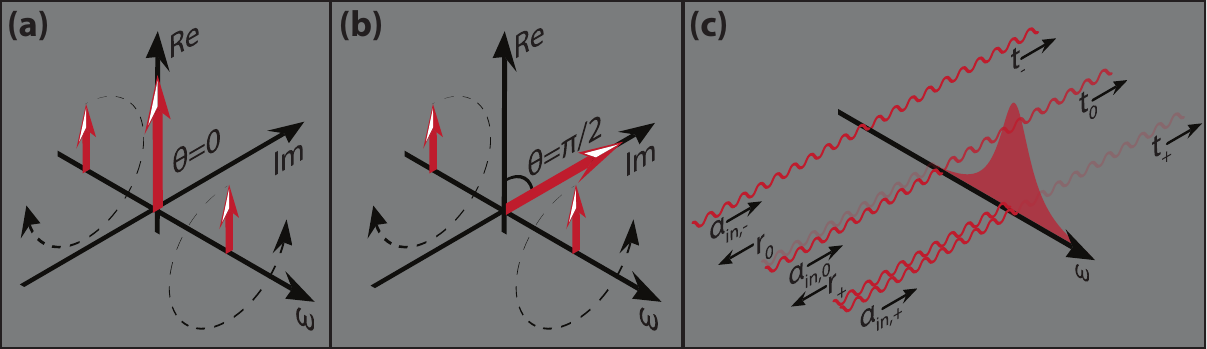}
\caption{Phasor illustration of amplitude and phase modulation, and expected cavity transmission and reflection for a modulated input field.  (a) Frequency components of pure amplitude modulation in a frame rotating at the carrier frequency. The two sidebands are placed at $\pm \omega_\text{m}$, where $\omega_\text{m}$ is the frequency of modulation. Here the modulation is in a direction parallel to the carrier tone. (b) Frequency components of a pure phase modulation. Unlike the case of amplitude modulation, the modulation occurs in a direction perpendicular to the carrier tone. (c) Illustration of the expected reflected and transmitted signals for the case of a red detuned pump laser modulated such that one sideband is near resonance with the cavity.}
\label{fig:eomPlot}
\end{figure}

The results demonstrated here require strong control fields and weak probe fields, which were generated through EOM modulation of the control fields. Due to the available equipment, a phase EOM was used for the mode near $\omega_\text{a}$ and an amplitude EOM for the mode near $\omega_\text{c}$. This leads to differences in the probe transduction as measured on the high speed PD.

For weak modulation ($\beta \ll 1$, where $\beta$ is the index of modulation), we can assume the output of the EOM has three distinct frequency components at $\omega,\, \omega\pm\omega_\text{m}$, where $\omega$ is the frequency of the carrier tone and $\omega_\text{m}$ is the frequency of modulation. For convenience we will work in a frame rotating with the carrier at $\omega$. The type of modulation can be inferred from the sum of the sideband components in the imaginary plane. For pure amplitude modulation they will oscillate parallel to the carrier with frequency $\omega_\text{m}$, whereas for pure phase modulation they will oscillate perpendicular to the carrier as illustrated in Supplementary Figure \ref{fig:eomPlot}(a,b). We note that since we are working in the rotating frame, we can choose the phase of the carrier out of convenience, as only the relative phase between the sidebands and the carrier tone influences the result.

In practice, the construction of amplitude EOMs is often such that the chirp is non-zero, which results in non-zero phase modulation of the outgoing field. With this in mind, we can write the transmission of the modulated field through the cavity as $\ket{\alpha_\text{out}} = t\ket{\alpha_\text{in}}$, where
\begin{equation}
t = \mathbf{diag}\{t_+, t_0, t_- \},\,
\ket{\alpha_\text{in}}= \alpha_\text{in}^0
\begin{pmatrix}
\frac{\beta}{2} e^{-i\omega_m t},
e^{i\theta},
\frac{\beta}{2} e^{i\omega_m t}
\end{pmatrix}^\mathbf{T}.
\end{equation}
In the previous expression, $t_+, t_0, t_-$ are the transmission coefficients at the upper sideband, carrier frequency, and lower sideband, respectively. The angle between the sidebands and the carrier is $\theta$, where $\theta = n\pi$ for a pure amplitude modulator and $\theta = \pi/2 \pm n\pi$ for a pure phase modulator, where $n$ is an integer.

The frequency components of the field transmitted through the cavity can be projected using the matrices
\begin{equation}
\bf{P_+} =
\begin{pmatrix}
0 & 1 & 0 \\
0 & 0 & 1 \\
0 & 0 & 0
\end{pmatrix},\,
\bf{P_0} =
\begin{pmatrix}
1 & 0 & 0 \\
0 & 1 & 0 \\
0 & 0 & 1
\end{pmatrix},\,
\bf{P_-} =
\begin{pmatrix}
0 & 0 & 0 \\
1 & 0 & 0 \\
0 & 1 & 0
\end{pmatrix}.
\end{equation}
Using the above expressions, we can write the signal measured on the PD up to a constant as $S = S_0 + S_1 + S_2$, where
\begin{align}
S_0 &=\bra{\alpha_\text{out}}P_0\ket{\alpha_\text{out}}, \\
S_1\cos (\omega_mt+\phi_1) &=\bra{\alpha_\text{out}}P_+ \ket{\alpha_\text{out}} + \bra{\alpha_\text{out}}P_-\ket{\alpha_\text{out}},\\
S_2\cos (2\omega_mt+\phi_2)&=\bra{\alpha_\text{out}}P_+^2 \ket{\alpha_\text{out}} + \bra{\alpha_\text{out}}P_-^2\ket{\alpha_\text{out}}.
\end{align}

By electronic filtering we isolate the $\mathcal{O}(\omega_\text{m})$ component of the signal, where
\begin{equation}
S_1 =\beta \left|t_0^* t_+ e^{-i\theta} + t_0 t_-^* e^{i\theta}\right|.
\end{equation}
To find the expected signal we use $\theta = 0.6093~\text{[Rad]}$, as measured directly from the OMIT spectra, and an agreement with the manufacturer's specifications. In this work the control laser is red detuned from a sideband-resolved cavity, as illustrated in Supplementary Figure \ref{fig:eomPlot}(c). In this case the lower sideband passes un--attenuated ($t_-=1$, $r_-=0$), and the control laser is approximately real $t_0 \approx t_0^*$. In this case, to first order in modulation angle, the expected signal is
\begin{equation}
S_1 \approx\beta t_0 \sqrt{(r_+-1-\cos(2\theta))^2 + \sin^2(2\theta)},
\end{equation}
where we have used the fact that $t_++r_+=1$. Using the chirp parameter \cite{ref:supp_yan2003mot} specified by the manufacturer for the amplitude EOM we calculate $\theta = 0.6093 $ [rad]. This value also agrees well to direct fits to the OMIT lineshapes.

In the case of pure phase modulation ($\theta = \pi/2 \pm n\pi$) this simply reduces to
\begin{equation}
S_1 \approx\beta t_0 |r_+|.
\end{equation}

\section*{Supplementary Note 3 - Data analysis}

In order to examine the bright and dark state coupling as shown in Fig.\ 3 in the main text, time-domain data was directly acquired on the DSA. For this dataset we digitally down mixed by the carrier frequency $\omega_\text{b}$, which allowed us to extract both the amplitude of the signal, and the phase relative to the carrier signal for modes $a$ and $c$. Due to chirp in the amplitude modulator, dispersion in the fiber, and difference in the optical path length of the two output arms of the WDM, a delay between the mode outputs was observed. To correct for this we fit the oscillating output of each mode to a sinusoidal function, and subtract the phase difference. Using this we are able to reconstruct the output of the dark and bright states.

\section*{Supplementary Note 4 - Double optomechanically induced transparency}

In this work, two optical modes $a$ and $c$ exhibit dispersive optomechanical coupling to the mechanical mode $b$. We denote the frequencies of these modes as $\omega_\text{a}$, $\omega_\text{c}$, and $\omega_\text{b}$, respectively, and the vacuum optomechanical coupling rates as $g_\text{a}$, and $g_\text{c}$. This is modelled by the Hamiltonian $\hat{H}=\hat{H}_0+\hat{H}_\text{int}$, where $\hat{H}_0$ describes the internal dynamics of each mode and $\hat{H}_\text{int}$ is the interaction Hamiltonian
\begin{align}
\hat{H}_0 & = \hbar \omega_\text{a} \hat{a}^\dagger \hat{a} + \hbar \omega_\text{b} \hat{b}^\dagger \hat{b} + \hbar \omega_\text{c} \hat{c}^\dagger \hat{c}, \label{eq:initialHamiltonian1} \\
\hat{H}_\text{int} & = -\hbar g_\text{a} \hat{a}^\dagger\hat{a} \left(\hat{b} + \hat{b}^\dagger \right) - \hbar g_\text{c} \hat{c}^\dagger\hat{c} \left(\hat{b} + \hat{b}^\dagger \right). \label{eq:initialHamiltonian2}
\end{align}
We describe the coupling between the optical modes and a waveguide using input-output theory
\begin{align}
\dot{\hat{a}} &= \frac{\1i}{\hbar} \left[\hat{H}, \hat{a}\right] - \frac{\kappa_\text{a}}{2}\hat{a} + \sqrt{\kappa_\text{a}^\text{ex}} \hat{a}_\text{in}, \label{eq:inputOutput1} \\
\dot{\hat{c}} &= \frac{\1i}{\hbar} \left[\hat{H}, \hat{c}\right] - \frac{\kappa_\text{c}}{2}\hat{c} + \sqrt{\kappa_\text{c}^\text{ex}} \hat{c}_\text{in}, \label{eq:inputOutput2}
\end{align}
where $\hat{a}_\text{in}$ and $\hat{c}_\text{in}$ are the input field operators for each optical mode, and $\kappa_\text{a}$, $\kappa_\text{c}$ and $\kappa_\text{a}^\text{ex}$, $\kappa_\text{c}^\text{ex}$ are the total energy decay and waveguide--cavity coupling rates of mode $a$ and $c$, respectively. Note that in this work the cavity is double-sided and consequently the cavity--waveguide coupling rate in each direction is $\kappa_\text{ex}/2$.

For all scenarios described in this work control lasers were red-detuned from the cavity modes whereas probe lasers were tuned near resonance. Although the modulators create multiple sidebands, the spectral selectivity of the cavity is such that only one sideband will contribute to the physics of the problem. This allows us to linearize about the control fields using the substitutions $\hat{a} \rightarrow \alpha_\text{a} + \hat{a}$, and $\hat{c} \rightarrow \alpha_\text{c} + \hat{c}$, where $\alpha_\text{a},\alpha_\text{c}$ are the classical control fields amplitudes, and $\hat{a}$, $\hat{c}$ now represent the cavity fluctuations near the probe frequencies. We also use similar substitutions for the input field amplitudes, such that $\hat{a}_\text{in}$ and $\hat{c}_\text{in}$ are the input probe field operators. Neglecting small order terms, and accounting for a static mechanical shift induced by constant radiation pressure, our interaction Hamiltonian becomes
\begin{equation}
\hat{H}_\text{int} = \ -\hbar g_\text{a} \left(\alpha_\text{a}\hat{a}^\dagger + \alpha^*_\text{a} \hat{a} \right) \left(\hat{b} + \hat{b}^\dagger \right)
-\hbar g_\text{c} \left(\alpha_\text{c}\hat{c}^\dagger + \alpha^*_\text{c} \hat{c} \right) \left(\hat{b} + \hat{b}^\dagger \right).
\label{eq:linearizedHamiltonian}
\end{equation}

We consider the case where the control lasers are red--detuned, with the probe fields on resonance such that $\Delta^\text{ctrl}_\text{i} = \omega^\text{ctrl}_\text{i}-\omega_\text{i} = -\omega_\text{b}$, and $\omega^{\text{probe}}_\text{i}-\omega^{\text{ctrl}}_\text{i} = \omega_\text{b}$, where $i =$ \{a,c\}. In this case, selecting only the resonant terms under the rotating wave approximation, the above expression simplifies to
\begin{equation}
\hat{H}_\text{int} = -\hbar\left(G_\text{a}\hat{a}^\dagger\hat{b} + G^*_\text{a} \hat{a} \hat{b}^\dagger +
G_\text{c}\hat{c}^\dagger\hat{b} + G^*_\text{c} \hat{c} \hat{b}^\dagger \right),
\label{eq:redHamiltonian}
\end{equation}
where $G_\text{a} = \alpha_\text{a} g_\text{a}$ and $G_\text{c} = \alpha_\text{c} g_\text{c}$. Transforming into frequency space, in a frame rotating with the control lasers, and making use of Supplementary Equations (\ref{eq:inputOutput1}-\ref{eq:inputOutput2}) and (\ref{eq:redHamiltonian}) we may solve for the mode operators using the set of coupled linear equations
\begin{align}
\begin{bmatrix}
\chi_\text{a}^{-1}(\omega) & -\1i G_\text{a} & 0 \\
-\1i G_\text{a}^* & \chi_\text{b}^{-1}(\omega) & -\1iG_\text{c}^* \\
0 & -\1i G_\text{c} & \chi_\text{c}^{-1}(\omega)
\end{bmatrix}
\begin{bmatrix}
\hat{a} \\
\hat{b} \\
\hat{c}
\end{bmatrix}
=
\begin{bmatrix}
\sqrt{\kappa_\text{a}^\text{ex}} \hat{a}_\text{in}\\
\sqrt{\gamma_\text{b}}\hat{b}_\text{in}\\
\sqrt{\kappa_\text{c}^\text{ex}} \hat{c}_\text{in}
\end{bmatrix}.
\end{align}

In the above we have written the cavity susceptibilities as $\chi_\text{a}^{-1}(\omega)=\kappa_\text{a}/2-\1i(\Delta_\text{a}+\omega)$, and $\chi_\text{c}^{-1}(\omega)=\kappa_\text{c}/2-\1i(\Delta_\text{c}+\omega)$, where, for notational cleanliness we have defined $\Delta_\text{a} = \Delta^\text{ctrl}_\text{a}$, and $\Delta_\text{c} = \Delta^\text{ctrl}_\text{c}$. We also define the mechanical susceptibility as $\chi_\text{b}^{-1}(\omega)=\gamma_\text{b}/2-\1i(-\omega_\text{b}+\omega)$, including a mechanical input field, $\hat{b}_\text{in}$, which can be used to model thermal contact with the environment.

From here the solutions become tractable if we make a change of basis to symmetric and antisymmetric combinations of the $a$ and $c$ modes which we refer to as the mechanically dark, $\zeta_\text{dk}$, and bright, $\zeta_\text{br}$, modes \cite{ref:supp_dong2012odm}
\begin{align}
\hat{\zeta}_\text{dk} &= \frac{ G_\text{c} \hat{a} - G_\text{a} \hat{c}}{\1i\overline{G}}, \\
\hat{\zeta}_\text{br} &= \frac{ G^*_\text{a} \hat{a} + G^*_\text{c} \hat{c} }{\overline{G}},
\end{align}
where
\begin{equation}
\overline{G} = \sqrt{|G_\text{a}|^2+|G_\text{c}|^2}.
\end{equation}

\noindent Assuming $\kappa_1 = \kappa_2 = \kappa$, and $\Delta_\text{a} = \Delta_\text{c} = \Delta$, we arrive at de--coupled equations of motion, which have the solutions
\begin{align}
\hat{\zeta}_\text{dk}
&=
\frac{1}{\kappa/2-\1i(\Delta+\omega)}\left( \frac{\sqrt{\kappa^\text{ex}_\text{a}}G_\text{c}\hat{a}_\text{in} - \sqrt{\kappa^\text{ex}_\text{c}}G_\text{a}\hat{c}_\text{in} }{\1i\overline{G}} \right),\\
\hat{b}
&=
\frac{1}{\gamma_\text{b}/2-\1i(-\omega_\text{b} + \omega) }\left(\sqrt{\gamma_\text{b}} \hat{b}_\text{in} + \1i\overline{G}\hat{\zeta}_\text{br} \right), \\
\hat{\zeta}_\text{br}
&=
\frac{1}{\kappa/2-\1i(\Delta+\omega)+ \frac{ \overline{G}^2 }{\gamma_\text{b}/2-\1i(-\omega_\text{b} + \omega) }} \left( \frac{\sqrt{\kappa^\text{ex}_\text{a}}G_\text{a}^*\hat{a}_\text{in} + \sqrt{\kappa^\text{ex}_\text{c}}G_\text{c}^*\hat{c}_\text{in} }{\overline{G}} + \frac{\1i \overline{G} \sqrt{\gamma_\text{b}} \hat{b}_\text{in}}{\gamma_\text{b}/2-\1i(-\omega_\text{b} + \omega) } \right).
\end{align}

To easily access the physics of the system we take $G_\text{a}=G_\text{c}=G$, and $\kappa_\text{a}^\text{ex}=\kappa_\text{c}^\text{ex}=\kappa_\text{ex}$ to simplify these expressions. We also ignore any input mechanical drive by setting $\hat{b}_\text{in}\rightarrow \ 0$. Finally, we assume classical probe fields of equal amplitude, $s_\text{in}$, and drive each modulator at the same frequency with phase difference $\phi$ by making the substitutions $\hat{a}_\text{in}\rightarrow s_\text{in}e^{i\phi/2}$ and $\hat{c}_\text{in}\rightarrow s_\text{in}e^{-i\phi/2}$. This results in the expressions,
\begin{align}
\zeta_\text{dk}
&=
\frac{\sqrt{2\kappa_\text{ex}} \sin(\phi/2) s_\text{in}}{\kappa/2-\1i(\Delta+\omega)}, \label{eqn:dk} \\
\zeta_\text{br}
&=
\frac{\sqrt{2\kappa_\text{ex}}\cos(\phi/2) s_\text{in}}{\kappa/2 - \1i(\Delta + \omega)+\frac{ 2 G^2 }{\gamma_\text{b}/2 - \1i(-\omega_\text{b} + \omega)}}.\label{eqn:br}
\end{align}

\section*{Supplementary Note 5 - Effect of mismatched parameters}

In the above expressions, we developed a model assuming idealized parameters. This resulted in mechanically bright and mechanically dark states which were decoupled from each other, and which could be isolated by adjusting the phase of the probe lasers. However, in any physical implementation of DOMIT, there will be mismatch between various parameters. In the following sections we study the effect of these mismatched parameters one by one.

\subsection*{Mismatched probe amplitudes}

Suppose that all parameters are matched according to the set of assumptions the led to Supplementary Equations (\ref{eqn:dk}) and (\ref{eqn:br}). We can include the effect of probe mismatch by instead making the substitutions $\hat{a}_\text{in}\rightarrow \left(s_\text{in} + \delta_\text{s} \right)e^{i\phi/2}$ and $\hat{c}_\text{in}\rightarrow \left(s_\text{in} - \delta_\text{s} \right) e^{-i\phi/2}$, where $s_\text{in}$ is the average probe power, and $2\delta_\text{s}$ is the difference in the probe powers. Proceeding as before, we find,
\begin{align}
\zeta_\text{dk}
&=
\frac{\sqrt{2\kappa_\text{ex}} \left(\sin(\phi/2) s_\text{in} -i\cos(\phi/2) \delta_\text{s} \right) }{\kappa/2-\1i(\Delta+\omega)}, \label{eqn:dkProbeMismatch} \\
\zeta_\text{br}
&=
\frac{\sqrt{2\kappa_\text{ex}} \left(\cos(\phi/2) s_\text{in} - i \sin(\phi/2) \delta_\text{s} \right) }{\kappa/2 - \1i(\Delta + \omega)+\frac{ 2 G^2 }{\gamma_\text{b}/2 - \1i(-\omega_\text{b} + \omega)}} . \label{eqn:brProbeMismatch}.
\end{align}
From the above expression, one can see that for $\delta_\text{s} \neq 0$ no choice of $\phi$ will enable complete isolation of the dark or bright state. This is further elucidated by calculating the dependence of the mode energy on $\phi$ and $\delta_\text{s}$ for constant input probe power,
\begin{align}
|\zeta_\text{dk}|^2 &\propto \frac{\left(|s_\text{in}|^2-|\delta_\text{s}|^2 \right)\sin^2(\phi/2) + |\delta_\text{s}|^2}{|s_\text{in}|^2+|\delta_\text{s}|^2 }, \\
|\zeta_\text{br}|^2 &\propto \frac{\left(|s_\text{in}|^2-|\delta_\text{s}|^2 \right)\cos^2(\phi/2) + |\delta_\text{s}|^2}{|s_\text{in}|^2+|\delta_\text{s}|^2 }.
\end{align}

\begin{figure}[ht]
\centering
\includegraphics[width=\linewidth]{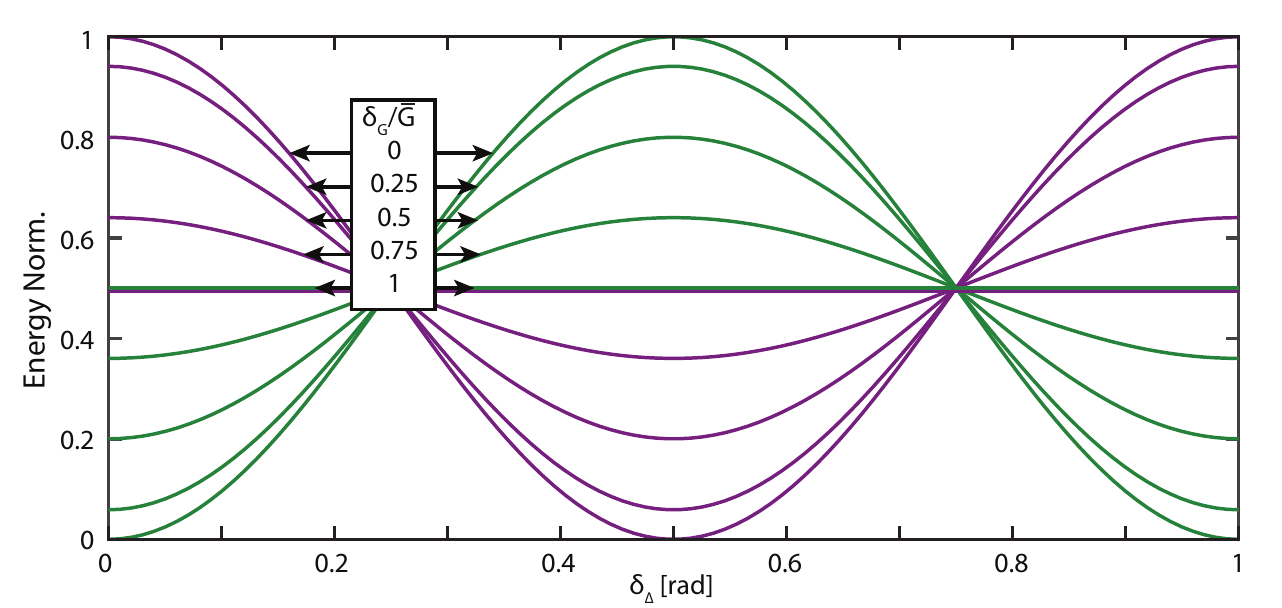}
\caption{Effect of imbalance in the probe powers, or optomechanical coupling rates on the amplitude of the bright and dark states. The amplitude of the dark (green) and bright (purple) state as a function of cavity-probe detuning, $\delta_\Delta$ is reduced as the mismatch in optomechanical coupling rate, $\delta_G$, increases.}
\label{fig:mismatchPlot}
\end{figure}

\subsection*{Mismatched optomechanical coupling}

The effect of mismatch in the optomechanical coupling will have similar effects to mismatch in the probe amplitudes. This can be included by making the substitutions $G_\text{a}\rightarrow G+ \delta_\text{G}$ and $G_\text{c}\rightarrow G - \delta_\text{G}$, where $G$ is the average optomechanical coupling rate, and $2\delta_\text{G}$ is the difference in optomechanical coupling rates. With these substitutions, the amplitudes of the mechanically bright and mechanically dark state are,
\begin{align}
\zeta_\text{dk}
&=
\frac{\sqrt{2\kappa_\text{ex}} s_\text{in} }{\kappa/2-\1i(\Delta+\omega)} \left(\frac{G \sin(\phi/2) + i\delta_\text{G}\cos(\phi/2)}{\overline{G}} \right), \\
\zeta_\text{br}
&=
\frac{\sqrt{2\kappa_\text{ex}} s_\text{in}}{\kappa/2 - \1i(\Delta + \omega)+\frac{ 2 G^2 }{\gamma_\text{b}/2 - \1i(-\omega_\text{b} + \omega)}}
\left(\frac{G \cos(\phi/2) s_\text{in} + i \delta_\text{G} \sin(\phi/2)}{\overline{G}} \right) ,
\end{align}
where $\overline{G} = \sqrt{G^2+\delta_\text{G}^2}$ in this case. Calculating the mode amplitudes, we find,
\begin{align}
|\zeta_\text{dk}|^2 &\propto \frac{\left(|G|^2-|\delta_\text{G}|^2 \right)\sin^2(\phi/2) + |\delta_\text{G}|^2}{|\overline{G}|^2}, \\
|\zeta_\text{br}|^2 &\propto \frac{\left(|G|^2-|\delta_\text{G}|^2 \right)\cos^2(\phi/2) + |\delta_\text{G}|^2}{|\overline{G}|^2}.
\end{align}

\subsection*{Mismatched frequency and damping}

Up until this point, we have found no direct coupling between the bright and dark mode. However, by detuning either our probe or pump lasers in equal and opposite directions, we can induce a coupling between these two modes. Furthermore, as we shall show, a mismatching the damping rates of the optical modes will also lead to a coupling. To see this, we make the substitutions $\Delta_\text{a} \rightarrow \Delta + \delta_\Delta$, $\Delta_\text{c} \rightarrow \Delta - \delta_\Delta$, $\kappa_\text{a} \rightarrow \kappa + \delta_\kappa$, and $\kappa_\text{c} \rightarrow \kappa - \delta_\kappa$. To clarify matters, we assume that the input mechanical is negligible ($\hat{b}_\text{in}\rightarrow \ 0$), and set $G_\text{a}=G_\text{c}=G$, and $\kappa_1 = \kappa_2 = \kappa$ in Eqn. (4), which gives
\begin{align}
\chi^{-1}(\omega)\hat{\zeta}_\text{dk} &= \frac{\sqrt{\kappa_\text{a}^\text{ex}}\hat{a}_\text{in}-\sqrt{\kappa_\text{c}^\text{ex}}\hat{c}_\text{in}}{\sqrt{2}\1i}+\left(\delta_\Delta + i\frac{\delta_\kappa}{2} \right)\hat{\zeta}_\text{br},\\
\chi^{-1}(\omega)\hat{\zeta}_\text{br} &= \frac{\sqrt{\kappa_\text{a}^\text{ex}}\hat{a}_\text{in}+\sqrt{\kappa_\text{c}^\text{ex}}\hat{c}_\text{in}}{\sqrt{2}}+\1i\overline{G}\hat{b}+\left(\delta_\Delta + i\frac{\delta_\kappa}{2} \right)\hat{\zeta}_\text{dk}.
\end{align}
\noindent where $\chi^{-1}(\omega)=\kappa/2-\1i(\Delta+\omega)$. From these expressions we see that there is coupling between bright and dark mode. For differences in frequency we have dispersive coupling, at a rate $\delta_\Delta$, whereas for differences in damping, we have dissipative coupling at a rate $\delta_\kappa/2$.

\section*{Supplementary Note 6 - Dark-Bright mode coupling}

In this section, we consider dissipative coupling due to detuning either our probe or pump lasers in equal and opposite directions. To describe this coupling in the time domain we first consider the intermodal coupling for the case $\delta_\Delta=0$, as illustrated in Supplementary Figure \ref{fig:s3}(a). Here, depending on the relative phase of the probe lasers, we arrive at a superposition of $\zeta_\text{br}$ and $\zeta_\text{dk}$ which is constant in time. In order for this process to remain stationary, we require interference to be between oscillations of the same frequency. From this, we can infer that by shifting the probe-cavity detuning by an amount $\delta_\Delta$, we cause interference to occur between differing frequencies, leading to beating between modes as illustrated in Supplementary Figure \ref{fig:s3}(b).

\begin{figure}[ht]
\centering
\includegraphics[width=\linewidth]{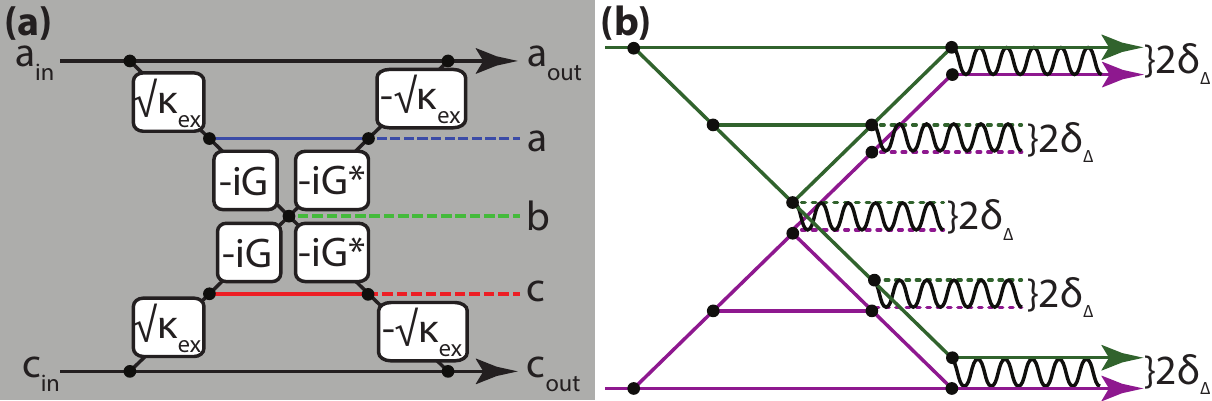}
\caption{Coupling the mechanically bright and dark states. (a) Intermodal coupling for the case of stationary DOMIT. Direction of propagation is left to right. (b) Effect of frequency shifting the control-probe detunings in opposite directions. Unlike the stationary case, we will observe a beating between all modes of the system.}
\label{fig:s3}
\end{figure}

To solve this, we divide each of modes into two frequency components at $\pm\delta_\Delta$ of our original frequency terms as illustrated in Supplementary Figure \ref{fig:s2}(b). Considering first the dark state in the time domain, and choosing to set $\phi=0$ for convenience, we find
\begin{equation}
\zeta_\text{dk}(t) = \zeta_\text{dk}^{(+)} + \zeta_\text{dk}^{(-)} =
\frac{e^{-\1i(\omega_\text{b}+\delta_\Delta)t}}{\kappa/2-\1i(\Delta+\omega_\text{b}+\delta_\Delta)}\sqrt{\frac{\kappa^\text{ex}}{2}}\frac{s_\text{in}}{\1i} +
\frac{e^{-\1i(\omega_\text{b}-\delta_\Delta)t}}{\kappa/2-\1i(\Delta+\omega_\text{b}-\delta_\Delta)}\sqrt{\frac{\kappa^\text{ex}}{2}}\frac{s_\text{in}}{\1i}.
\end{equation}
In a similar manner, we find that the bright state may be written as
\begin{align}
\zeta_\text{br}(t) =&\ \zeta_\text{br}^{(+)} + \zeta_\text{br}^{(-)}\nonumber\\
=&\
\frac{e^{-\1i(\omega_\text{b}+\delta_\Delta)t}}{\kappa/2-\1i(\Delta+\omega_\text{b}+\delta_\Delta)+\frac{|\overline{G}|^2}{\gamma_\text{b}/2-\1i(-\omega_\text{b}+\omega_\text{b}+\delta_\Delta)}}\sqrt{\frac{\kappa^\text{ex}}{2}}s_\text{in} \nonumber\\
&-\frac{e^{-\1i(\omega_\text{b}-\delta_\Delta)t}}{\kappa/2-\1i(\Delta+\omega_\text{b}-\delta_\Delta)+\frac{|\overline{G}|^2}{\gamma_\text{b}/2-\1i(-\omega_\text{b}+\omega_\text{b}-\delta_\Delta)}}\sqrt{\frac{\kappa^\text{ex}}{2}}s_\text{in}.
\end{align}

Setting $\Delta=-\omega_\text{b}$, and assuming $\delta \ll \kappa$, we find
\begin{align}
\zeta_\text{dk}(t) =& \frac{2\sqrt{\kappa_\text{ex}}\sin(\delta_\Delta t)e^{-\1i\omega_\text{b}t}s_\text{in}}{\kappa},\\
\zeta_\text{br}(t) =& \frac{2\sqrt{\kappa_\text{ex}}\cos(\delta_\Delta t)e^{-\1i\omega_\text{b}t}s_\text{in}}{\kappa(1+\frac{\overline{C}}{1+4(\delta_\Delta/\gamma_\text{b})^2})}.
\end{align}
This gives us the output fields as
\begin{align}
a(t) &= \frac{e^{-\1i\omega_\text{b}t}}{\sqrt{2}} \left(\1i\sin(\delta_\Delta t) \zeta_\text{dk}(0) + \cos(\delta_\Delta t) \zeta_\text{br}(0) \right), \\
c(t) &= \frac{e^{-\1i\omega_\text{b}t}}{\sqrt{2}} \left(-\1i\sin(\delta_\Delta t) \zeta_\text{dk}(0) + \cos(\delta_\Delta t) \zeta_\text{br} (0) \right).
\end{align}
We note that near resonance, the amplitudes $\zeta_\text{dk}$ and $\zeta_\text{br}$ approach those calculated for steady state. For cases where we detuned away from the DOMIT transparency ($\delta \gg \gamma_\text{b}$), or for small cooperativities, the amplitudes of the dark and light state approach each other, and the visibility of oscillations goes to zero. Explicitly, the resonance contrast of the oscillations is found to be
\begin{equation}
V(\delta) = 1-\sqrt{\frac{\gamma_\text{b}^2+4\delta_\Delta^2}{\gamma_\text{b}^2(1+\overline{C})^2+4\delta_\Delta^2}}.
\end{equation}

\section*{Supplementary Note 7 - Switching}
Although we previously found solutions in the frequency domain, it is instructive to reconsider the equations of motion in the time domain
\begin{align}
\dot{\hat{a}} = & \left(\1i \Delta_\text{a} - \kappa_\text{a}/2 \right)\hat{a} + \1iG_\text{a}\hat{b} + \sqrt{\kappa_\text{a}^\text{ex}} \hat{a}_\text{in}, \\
\dot{\hat{b}} = & \left(-\1i \omega_\text{b} - \gamma_\text{b}/2 \right)\hat{b} + \1iG^*_\text{a}\hat{a} + \1iG^*_\text{c}\hat{c}+\sqrt{\gamma_\text{b}}\hat{b}_\text{in}, \\
\dot{\hat{c}} = & \left(\1i \Delta_\text{c} - \kappa_\text{c}/2 \right)\hat{c} + \1iG_\text{c}\hat{b} + \sqrt{\kappa_\text{c}^\text{ex}} \hat{c}_\text{in}.
\end{align}

For the devices used in our experiment, the decay rate of our optics is much faster than our mechanics ($\kappa_\text{a}, \kappa_\text{c}\gg\gamma$). With this in mind we can use adiabatic elimination, and set $\dot{\hat{a}}=0, \dot{\hat{c}}=0$, and solve for the mechanics as

\begin{equation}
\dot{\hat{b}} = \left(-\1i \omega_\text{b} - \gamma_\text{b}/2 +\frac{\left|G_\text{a}\right|^2}{\1i\Delta_\text{a}-\kappa_\text{a}/2} +\frac{\left|G_\text{c}\right|^2}{\1i\Delta_\text{c}-\kappa_\text{c}/2} \right)\hat{b} -\frac{\1iG_\text{a}^*\sqrt{\kappa_\text{a}^\text{ex}} \hat{a}_\text{in}}{\1i \Delta_\text{a} - \kappa_\text{a}/2} -\frac{\1iG_\text{c}^*\sqrt{\kappa_\text{c}^\text{ex}} \hat{c}_\text{in}}{\1i \Delta_\text{c} - \kappa_\text{c}/2}+\sqrt{\gamma_\text{b}}\hat{b}_\text{in}.
\end{equation}

Using this expression we find that for $\Delta_\text{a} = \Delta_\text{c} = 0$, as in the experiment
\begin{equation}
\dot{\hat{b}} = - \left(\1i \omega_\text{b} + \tau^{-1}\right)\hat{b} + \frac{2\1i G_\text{a} \sqrt{\kappa_\text{a}^\text{ex}}\hat{a}_\text{in}}{\kappa_\text{a}} + \frac{2\1i G_\text{c} \sqrt{\kappa_\text{c}^\text{ex}}\hat{c}_\text{in}}{\kappa_\text{c}}+\sqrt{\gamma_\text{b}}\hat{b}_\text{in}.
\end{equation}
This gives the switching speed as $\tau^{-1}=\frac{\gamma_\text{b}}{2}\left(1 + C_\text{a} + C_\text{c} \right)$, where $C_\text{j} = 4G_\text{j}^2/\kappa_\text{j}\gamma_\text{b}$ is the optomechanical cooperativity and $ j = \{\text{a},\text{c}\}$.

Assuming travelling wave singlet modes, the transmission amplitudes through the switch are
\begin{align}
t_\text{br} = \frac{2\kappa_\text{ex}}{\kappa} \frac{1}{1+\overline{C}}-1,\\
t_\text{dk} = \frac{2\kappa_\text{ex}}{\kappa}-1.
\end{align}

\section*{Supplementary Note 8 - N-mode solution}
\subsection*{Mechanically bright and dark states}
The interference between probe fields in two optical modes can in principle be extended to any number of optical modes. Suppose we have a system where $N$ optical modes are dissipatively coupled to a single mechanical mode. We label the creation and annihilation operators associated with these optical modes as $\hat{a}_n^\dagger, \hat{a}_n$ where our index runs from zero to $N-1$ and the mechanical mode creation and annihilation operators as is $\hat{b}^\dagger, \hat{b}$. We assume red detuned pumps, and that the system is sideband resolved. Our interaction Hamiltonian is:
\begin{equation}\label{eq:nModeHamiltonian}
\hat{H}_\text{int} = -\hbar \sum_{n=0}^{N-1} \left(G_n \hat{a}_n^\dagger \hat{b} + G^*_n \hat{a}_n \hat{b}^\dagger \right).
\end{equation}

Using the same set of assumptions as the DOMIT section above, we can use Supplementary Equation (\ref{eq:nModeHamiltonian}) and the input-output formalism to write a set of $N+1$ coupled equations,
\begin{align}
\dot{\hat{a}}_n = & \left(i \Delta_n -\frac{\kappa_n}{2} \right) \hat{a}_n + iG_n \hat{b} + \sqrt{\kappa_n^\text{ex}} \hat{a}_n^\text{in}, \\
\dot{\hat{b}} = & \left(-i\omega_b - \frac{\Gamma_b}{2} \right) \hat{b} + i\sum_{n=0}^{N-1} G_n^* \hat{a}_n + \sqrt{\gamma_b} \hat{b}^\text{in},
\end{align}
where we have indexed the decay rates for each mode to account for the possibility of mismatched decay rates.

This set of equations may be solved in a similar manner to DOMIT by seeking for a new basis where one mode is maximally coupled to the mechanics. To do so, we assume $\kappa_n = \kappa$, $\Delta_n = \Delta$, and $G_n = G$. Inspection of Supplementary Equation (\ref{eq:nModeHamiltonian}), reveals that the mechanics couples to the summation of all optical modes. Designating this as our mechanically bright mode, it then remains to construct a set of $N-1$ orthogonal modes. This may be achieved by applying a discrete Fourier transform to the optical modes,
\begin{equation}
\hat{\zeta}_m = \frac{1}{\sqrt{N}}\sum_{n=0}^{N-1} \hat{a}_n e^{-\frac{2\pi i}{N} nm}.
\end{equation}
Here $\zeta_0$ is the mechanically bright mode, and all others are mechanically dark modes. The optical input to the cavity is defined in the same manner,
\begin{equation}
\hat{\zeta}_m^\text{in} = \frac{1}{\sqrt{N}}\sum_{n=0}^{N-1} \hat{a}_n^\text{in} e^{-\frac{2\pi i}{N} nm}.
\end{equation}

In this basis, our equations take on the simpler form,
\begin{align}
\dot{\hat{\zeta}}_0 = & \left( i \Delta -\frac{\kappa}{2}\right)\hat{\zeta}_0 + i \overline{G} \hat{b} + \sqrt{\kappa^\text{ex}}\hat{\zeta}_0^\text{in},\\
\dot{\hat{b}} = & \left( -i \omega_b -\frac{\gamma_b}{2}\right) \hat{b} + i \overline{G} \hat{\zeta}_0 + \sqrt{\gamma_b} \hat{b}^\text{in},\\
\dot{\hat{\zeta}}_{m} = & \left( i \Delta -\frac{\kappa}{2}\right) \hat{\zeta}_{m} + \sqrt{\kappa^\text{ex}} \hat{\zeta}_m^\text{in}.
\end{align}
Where in the above the m index in the $\zeta_m$ equations runs from 1 to $N$, and $\overline{G} = \sqrt{N}G$.

These equations can be solved by transforming into frequency space,
\begin{align}
\hat{\zeta}_0(\omega) &= \frac{1}{\kappa/2 - i(\omega+\Delta)+\frac{N|G|^2}{\gamma_b/2-i(\omega-\omega_b)}} \left(\sqrt{\kappa^\text{ex}}\hat{\zeta}_0^\text{in} + \frac{i\sqrt{\gamma_\text{b}}\overline{G}\hat{b}^\text{in}}{\gamma_b/2-i(\omega-\omega_b)}\right), \\
\hat{b}(\omega) &= \frac{1}{\gamma/2 - i(\omega-\omega_\text{b})+\frac{N|G|^2}{\kappa/2-i(\omega+\Delta)}} \left(\sqrt{\gamma_\text{b}}\hat{b}^\text{in} + \frac{i\sqrt{\kappa^\text{ex}}\,\overline{G} \hat{\zeta}_0^\text{in}}{\kappa/2-i(\omega+\Delta)}\right), \\
\zeta_m(\omega) &= \frac{\sqrt{\kappa^\text{ex}} \hat{\zeta}_m^\text{in}}{\kappa/2 - i(\omega+\Delta)}.
\end{align}

To gain insight into these expressions, we consider the on resonance ($\Delta = -\omega_\text{b},\omega = \omega_\text{b}$). This yields the expressions
\begin{align}
\hat{\zeta}_0(\omega) &= \frac{2/\kappa}{1+\overline{C}} \left(\sqrt{\kappa^\text{ex}}\hat{\zeta}_0^\text{in} + \frac{i2\overline{G}\hat{b}^\text{in}}{\sqrt{\gamma_\text{b}}}\right), \\
\hat{b}(\omega) &= \frac{2/\gamma}{1+\overline{C}} \left(\sqrt{\gamma_\text{b}}\hat{\hat{b}}^\text{in} + \frac{i2\overline{G} \zeta_0^\text{in} }{\kappa/\sqrt{\kappa^\text{ex}}}\right), \\
\zeta_m(\omega) &= 2\sqrt{\kappa^\text{ex}}/\kappa \hat{\zeta}_m^\text{in}.
\end{align}
From these expressions it can be seen that optomechanical coupling is only present between the mechanics and the mechanically bright mode. All other optical modes will see a bare cavity response. Initially, as the optomechanical coupling is increased, the degree of exchange between the bright mode and the mechanics will also increase. This situation has some resemblance to an add-drop filter. For very large cooperativities, both the mechanics and the bright mode will suppressed, and the system acts as notch filter for these modes.

Interestingly, this general case subsumes many well studied optomechanical effects. For example, in the case of a single optical mode, only the bright state can exist. In this case the filtering effect describes OMIT, where the occupation of the optical cavity is suppressed. In the case of two optical modes, as discussed previously in this paper, both a mechanically bright and a dark mode may exist. In this case, selection of the mechanically dark mode can lead to coupling between different colours of input light, while avoiding decoherence due to the mechanics. The analysis here indicates that for larger dimensions, there will always exist $N-1$ such mechanically dark modes, which can avoid decoherence from the mechanics.

\subsection*{Outputs}
Here we will calculate the output in the more physical basis of the individual optical modes. If we assume that $b^\text{in}$ can be neglected, we can write out solutions in the simple form $\zeta_0 = \eta_0 \zeta_0^\text{in}$ for $m=0$, and $\zeta_m = \eta_1 \zeta_m^\text{in}$ otherwise. Next we would like to return to our original basis in order to calculate the transmission at the physical ports. To do this, we use the inverse discrete Fourier transform, defined as
\begin{equation}
\hat{a}_n = \frac{1}{\sqrt{N}} \sum_{m=0}^{N-1}\hat{\zeta}_m e^{\frac{2\pi i}{N} nm}.
\end{equation}

Placing our solutions into this expression we find
\begin{align}
a_n = & \frac{\eta_0}{\sqrt{N}}\zeta_0^\text{in} + \frac{\eta_1}{\sqrt{N}} \sum_{m=1}^{N-1} \zeta_m^\text{in} e^{\frac{2\pi i}{N} nm}, \nonumber\\
= & \frac{\eta_0-\eta_1}{N} \sum_{n'=0}^{N-1} a_{n'}^{in} + \frac{\eta_1}{N}\sum_{n'=0}^{N-1} \left( \sum_{m=0}^{N-1} e^{\frac{-2\pi i}{N} m(n'-n)} \right)a_{n'}^\text{in}, \nonumber\\
= & \frac{\eta_0-\eta_1}{N} \sum_{n'=0}^{N-1} a_{n'}^{in} + \frac{\eta_1}{N}\sum_{n'=0}^{N-1} N \delta_{n'n} a_{n'}^\text{in}, \nonumber\\
= & \frac{\eta_0-\eta_1}{N} \sum_{n'=0}^{N-1} a_{n'}^{in} + \eta_1 a_n.
\end{align}

If we take the on-resonance case, $\eta_0 = \frac{2\sqrt{\kappa^\text{ex}}}{\kappa}$ and $\eta_1 = \frac{2\sqrt{\kappa^\text{ex}}}{\kappa}\frac{1}{1+\overline{C}}$ where $\overline{C} = NC = \frac{4N|G|^2}{\kappa \Gamma_b}$. This allows us to write,
\begin{align}
a_n = &\frac{2\sqrt{\kappa^\text{ex}}}{\kappa} \left(a_n^\text{in} - \frac{\overline{C}}{N(1+\overline{C})}\sum_{n'=0}^{N-1}a_{n'}^\text{in} \right), \\
a_n^\text{out} = & \left( 1 - \frac{2\kappa^\text{ex}}{\kappa} a_n^\text{in}\right) + \frac{2\kappa^\text{ex}}{\kappa}\frac{\overline{C}}{N(1+\overline{C})}\sum_{n'=0}^{N-1}a_{n'}^\text{in}.
\end{align}

For the case of critical coupling, and large cooperativity,
\begin{equation}
a_n^\text{out} = - a_n^\text{in} + \frac{2}{N} \sum_{n'=0}^{N-1}a_{n'}^\text{in}.
\end{equation}

This can be cast in matrix form as,
\begin{equation}
\begin{pmatrix}
a_0^\text{out} \\
a_1^\text{out} \\
\vdots \\
a_{N-1}^\text{out}
\end{pmatrix}
= \frac{1}{N}
\begin{pmatrix}
2-N & 2 & \cdots & 2 \\
2 & 2-N & \cdots & 2 \\
\vdots & \vdots & \ddots & \vdots \\
2 & 2 & \cdots & 2-N
\end{pmatrix}
\begin{pmatrix}
a_0^\text{in} \\
a_1^\text{in} \\
\vdots \\
a_{N-1}^\text{in}
\end{pmatrix}.
\end{equation}
This indicates that complete conversion from one colour to another is only possible for the case $N=2$.


\begin{thebibliography}{10}
\expandafter\ifx\csname url\endcsname\relax
  \def\url#1{\texttt{#1}}\fi
\expandafter\ifx\csname urlprefix\endcsname\relax\def\urlprefix{URL }\fi
\providecommand{\bibinfo}[2]{#2}
\providecommand{\eprint}[2][]{\url{#2}}

\bibitem{ref:LIGO2017gw170104}
\bibinfo{author}{Abbott, B.} \emph{et~al.}
\newblock \bibinfo{title}{{GW}170104: observation of a 50-solar-mass binary
  black hole coalescence at redshift 0.2}.
\newblock \emph{\bibinfo{journal}{Phys. Rev. Lett.}}
  \textbf{\bibinfo{volume}{118}}, \bibinfo{pages}{221101}
  (\bibinfo{year}{2017}).

\bibitem{ref:yariv2006poe}
\bibinfo{author}{Yariv, A.} \& \bibinfo{author}{Yeh, P.}
\newblock \emph{\bibinfo{title}{Photonics: Optical Electronics in Modern
  Communications}} (\bibinfo{publisher}{Oxford, New York},
  \bibinfo{year}{2006}).

\bibitem{ref:liu2004hss}
\bibinfo{author}{Liu, A.} \emph{et~al.}
\newblock \bibinfo{title}{A high-speed silicon optical modulator based on a
  metal--oxide--semiconductor capacitor}.
\newblock \emph{\bibinfo{journal}{Nature}} \textbf{\bibinfo{volume}{427}},
  \bibinfo{pages}{615--618} (\bibinfo{year}{2004}).

\bibitem{ref:obrien2009pqt}
\bibinfo{author}{O'Brien, J.}, \bibinfo{author}{Furusawa, A.} \&
  \bibinfo{author}{Vu\v{c}kovi\'{c}, J.}
\newblock \bibinfo{title}{Photonic quantum technologies}.
\newblock \emph{\bibinfo{journal}{Nat. Photonics}}
  \textbf{\bibinfo{volume}{3}}, \bibinfo{pages}{687--695}
  (\bibinfo{year}{2009}).

\bibitem{ref:clemmen2016ramsey}
\bibinfo{author}{Clemmen, S.}, \bibinfo{author}{Farsi, A.},
  \bibinfo{author}{Ramelow, S.} \& \bibinfo{author}{Gaeta, A.~L.}
\newblock \bibinfo{title}{Ramsey interference with single photons}.
\newblock \emph{\bibinfo{journal}{Phys. Rev. Lett.}}
  \textbf{\bibinfo{volume}{117}}, \bibinfo{pages}{223601}
  (\bibinfo{year}{2016}).

\bibitem{ref:lu2018quantum}
\bibinfo{author}{Lu, H.-H.} \emph{et~al.}
\newblock \bibinfo{title}{Quantum interference and correlation control of
  frequency-bin qubits}.
\newblock \emph{\bibinfo{journal}{Optica}} \textbf{\bibinfo{volume}{5}},
  \bibinfo{pages}{1455--1460} (\bibinfo{year}{2018}).

\bibitem{ref:kues2017chip}
\bibinfo{author}{Kues, M.} \emph{et~al.}
\newblock \bibinfo{title}{On-chip generation of high-dimensional entangled
  quantum states and their coherent control}.
\newblock \emph{\bibinfo{journal}{Nature}} \textbf{\bibinfo{volume}{546}},
  \bibinfo{pages}{622--626} (\bibinfo{year}{2017}).

\bibitem{ref:kurizki2015quantum}
\bibinfo{author}{Kurizki, G.} \emph{et~al.}
\newblock \bibinfo{title}{Quantum technologies with hybrid systems}.
\newblock \emph{\bibinfo{journal}{Proceedings of the National Academy of
  Sciences}} \textbf{\bibinfo{volume}{112}}, \bibinfo{pages}{3866--3873}
  (\bibinfo{year}{2015}).

\bibitem{ref:harris1998photon}
\bibinfo{author}{Harris, S.} \& \bibinfo{author}{Yamamoto, Y.}
\newblock \bibinfo{title}{Photon switching by quantum interference}.
\newblock \emph{\bibinfo{journal}{Phys. Rev. Lett.}}
  \textbf{\bibinfo{volume}{81}}, \bibinfo{pages}{3611--3614}
  (\bibinfo{year}{1998}).

\bibitem{ref:chang2014quantum}
\bibinfo{author}{Chang, D.~E.}, \bibinfo{author}{Vuleti{\'c}, V.} \&
  \bibinfo{author}{Lukin, M.~D.}
\newblock \bibinfo{title}{Quantum nonlinear optics--photon by photon}.
\newblock \emph{\bibinfo{journal}{Nat. Photonics}}
  \textbf{\bibinfo{volume}{8}}, \bibinfo{pages}{685--694}
  (\bibinfo{year}{2014}).

\bibitem{ref:mcguinness2010quantum}
\bibinfo{author}{McGuinness, H.~J.}, \bibinfo{author}{Raymer, M.~G.},
  \bibinfo{author}{McKinstrie, C.~J.} \& \bibinfo{author}{Radic, S.}
\newblock \bibinfo{title}{Quantum frequency translation of single-photon states
  in a photonic crystal fiber}.
\newblock \emph{\bibinfo{journal}{Phys. Rev. Lett.}}
  \textbf{\bibinfo{volume}{105}}, \bibinfo{pages}{093604}
  (\bibinfo{year}{2010}).

\bibitem{ref:kobayashi2016frequency}
\bibinfo{author}{Kobayashi, T.} \emph{et~al.}
\newblock \bibinfo{title}{Frequency-domain {H}ong--{O}u--{M}andel
  interference}.
\newblock \emph{\bibinfo{journal}{Nat. Photonics}}
  \textbf{\bibinfo{volume}{10}}, \bibinfo{pages}{441--444}
  (\bibinfo{year}{2016}).

\bibitem{ref:li2019tunable}
\bibinfo{author}{Li, Q.} \emph{et~al.}
\newblock \bibinfo{title}{Tunable quantum beat of single photons enabled by
  nonlinear nanophotonics}.
\newblock \emph{\bibinfo{journal}{Phys. Rev. Appl.}}
  \textbf{\bibinfo{volume}{12}}, \bibinfo{pages}{054054}
  (\bibinfo{year}{2019}).

\bibitem{ref:Qu2019CBD}
\bibinfo{author}{Qu, L.-Y.} \emph{et~al.}
\newblock \bibinfo{title}{Color erasure detectors enable chromatic
  interferometry}.
\newblock \emph{\bibinfo{journal}{Phys. Rev. Lett.}}
  \textbf{\bibinfo{volume}{123}}, \bibinfo{pages}{243601}
  (\bibinfo{year}{2019}).

\bibitem{ref:aspelmeyer2014co}
\bibinfo{author}{Aspelmeyer, M.}, \bibinfo{author}{Kippenberg, T.~J.} \&
  \bibinfo{author}{Marquardt, F.}
\newblock \bibinfo{title}{Cavity optomechanics}.
\newblock \emph{\bibinfo{journal}{Rev. Mod. Phys.}}
  \textbf{\bibinfo{volume}{86}}, \bibinfo{pages}{1391--1452}
  (\bibinfo{year}{2014}).

\bibitem{ref:safavi2011eit}
\bibinfo{author}{Safavi-Naeini, A.~H.} \emph{et~al.}
\newblock \bibinfo{title}{Electromagnetically induced transparency and slow
  light with optomechanics}.
\newblock \emph{\bibinfo{journal}{Nature}} \textbf{\bibinfo{volume}{472}},
  \bibinfo{pages}{69--73} (\bibinfo{year}{2011}).

\bibitem{ref:weis2010oit}
\bibinfo{author}{Weis, S.} \emph{et~al.}
\newblock \bibinfo{title}{Optomechanically induced transparency}.
\newblock \emph{\bibinfo{journal}{Science}} \textbf{\bibinfo{volume}{330}},
  \bibinfo{pages}{1520--1523} (\bibinfo{year}{2010}).

\bibitem{ref:fiore2013ops}
\bibinfo{author}{Fiore, V.}, \bibinfo{author}{Dong, C.},
  \bibinfo{author}{Kuzyk, M.~C.} \& \bibinfo{author}{Wang, H.}
\newblock \bibinfo{title}{Optomechanical light storage in a silica
  microresonator}.
\newblock \emph{\bibinfo{journal}{Phys. Rev. A}} \textbf{\bibinfo{volume}{87}},
  \bibinfo{pages}{023812} (\bibinfo{year}{2013}).

\bibitem{ref:hill2012cow}
\bibinfo{author}{Hill, J.~T.}, \bibinfo{author}{Safavi-Naeini, A.~H.},
  \bibinfo{author}{Chan, J.} \& \bibinfo{author}{Painter, O.}
\newblock \bibinfo{title}{Coherent optical wavelength conversion via cavity
  optomechanics}.
\newblock \emph{\bibinfo{journal}{Nat. Commun.}} \textbf{\bibinfo{volume}{3}},
  \bibinfo{pages}{1196} (\bibinfo{year}{2012}).

\bibitem{ref:dong2012ODM}
\bibinfo{author}{Dong, C.}, \bibinfo{author}{Fiore, V.},
  \bibinfo{author}{Kuzyk, M.~C.} \& \bibinfo{author}{Wang, H.}
\newblock \bibinfo{title}{Optomechanical dark mode}.
\newblock \emph{\bibinfo{journal}{Science}} \textbf{\bibinfo{volume}{338}},
  \bibinfo{pages}{1609--1613} (\bibinfo{year}{2012}).

\bibitem{ref:liu2013eit}
\bibinfo{author}{Liu, Y.}, \bibinfo{author}{Davan{\c{c}}o, M.},
  \bibinfo{author}{Aksyuk, V.} \& \bibinfo{author}{Srinivasan, K.}
\newblock \bibinfo{title}{Electromagnetically induced transparency and wideband
  wavelength conversion in silicon nitride microdisk optomechanical
  resonators}.
\newblock \emph{\bibinfo{journal}{Phys. Rev. Lett.}}
  \textbf{\bibinfo{volume}{110}}, \bibinfo{pages}{223603}
  (\bibinfo{year}{2013}).

\bibitem{ref:ockeloen2016LNA}
\bibinfo{author}{Ockeloen-Korppi, C.} \emph{et~al.}
\newblock \bibinfo{title}{Low-noise amplification and frequency conversion with
  a multiport microwave optomechanical device}.
\newblock \emph{\bibinfo{journal}{Phys. Rev. X}} \textbf{\bibinfo{volume}{6}},
  \bibinfo{pages}{041024} (\bibinfo{year}{2016}).

\bibitem{ref:barzanjeh2018ser}
\bibinfo{author}{Barzanjeh, S.} \emph{et~al.}
\newblock \bibinfo{title}{Stationary entangled radiation from micromechanical
  motion}.
\newblock \emph{\bibinfo{journal}{Nature}} \textbf{\bibinfo{volume}{570}},
  \bibinfo{pages}{480--483} (\bibinfo{year}{2019}).

\bibitem{ref:balram2016ccr}
\bibinfo{author}{Balram, K.~C.}, \bibinfo{author}{Davan{\c{c}}o, M.~I.},
  \bibinfo{author}{Song, J.~D.} \& \bibinfo{author}{Srinivasan, K.}
\newblock \bibinfo{title}{Coherent coupling between radiofrequency, optical and
  acoustic waves in piezo-optomechanical circuits}.
\newblock \emph{\bibinfo{journal}{Nat. Photonics}}
  \textbf{\bibinfo{volume}{10}}, \bibinfo{pages}{346--352}
  (\bibinfo{year}{2016}).

\bibitem{ref:korsunsky1999phase}
\bibinfo{author}{Korsunsky, E.~A.}, \bibinfo{author}{Leinfellner, N.},
  \bibinfo{author}{Huss, A.}, \bibinfo{author}{Baluschev, S.} \&
  \bibinfo{author}{Windholz, L.}
\newblock \bibinfo{title}{Phase-dependent electromagnetically induced
  transparency}.
\newblock \emph{\bibinfo{journal}{Phys. Rev. A}} \textbf{\bibinfo{volume}{59}},
  \bibinfo{pages}{2302--2305} (\bibinfo{year}{1999}).

\bibitem{ref:kang2006pcl}
\bibinfo{author}{Kang, H.}, \bibinfo{author}{Hernandez, G.},
  \bibinfo{author}{Zhang, J.} \& \bibinfo{author}{Zhu, Y.}
\newblock \bibinfo{title}{Phase-controlled light switching at low light
  levels}.
\newblock \emph{\bibinfo{journal}{Phys. Rev. A}} \textbf{\bibinfo{volume}{73}},
  \bibinfo{pages}{011802} (\bibinfo{year}{2006}).

\bibitem{ref:kang2011pcs}
\bibinfo{author}{Kang, H.}, \bibinfo{author}{Kim, B.}, \bibinfo{author}{Park,
  Y.~H.}, \bibinfo{author}{Oh, C.-H.} \& \bibinfo{author}{Lee, I.~W.}
\newblock \bibinfo{title}{Phase-controlled switching by interference between
  incoherent fields in a double-$\lambda$ system}.
\newblock \emph{\bibinfo{journal}{Opt. Express}} \textbf{\bibinfo{volume}{19}},
  \bibinfo{pages}{4113--4120} (\bibinfo{year}{2011}).

\bibitem{ref:Kim2015ALS}
\bibinfo{author}{Kim, B.}, \bibinfo{author}{Sohn, B.-U.},
  \bibinfo{author}{Shin, W.}, \bibinfo{author}{Ko, D.-K.} \&
  \bibinfo{author}{Kang, H.}
\newblock \bibinfo{title}{Anticorrelated light switching in an optical loop
  system}.
\newblock \emph{\bibinfo{journal}{J. Opt. Soc. Am. B}}
  \textbf{\bibinfo{volume}{32}}, \bibinfo{pages}{227--231}
  (\bibinfo{year}{2015}).

\bibitem{ref:xu2016topological}
\bibinfo{author}{Xu, H.}, \bibinfo{author}{Mason, D.}, \bibinfo{author}{Jiang,
  L.} \& \bibinfo{author}{Harris, J.}
\newblock \bibinfo{title}{Topological energy transfer in an optomechanical
  system with exceptional points}.
\newblock \emph{\bibinfo{journal}{Nature}} \textbf{\bibinfo{volume}{537}},
  \bibinfo{pages}{80--83} (\bibinfo{year}{2016}).

\bibitem{ref:Kuzyk2017MultiInt}
\bibinfo{author}{Kuzyk, M.~C.} \& \bibinfo{author}{Wang, H.}
\newblock \bibinfo{title}{Controlling multimode optomechanical interactions via
  interference}.
\newblock \emph{\bibinfo{journal}{Phys. Rev. A}} \textbf{\bibinfo{volume}{96}},
  \bibinfo{pages}{023860} (\bibinfo{year}{2017}).

\bibitem{ref:zhang2005aox}
\bibinfo{author}{Zhang, M.}, \bibinfo{author}{Wang, L.} \& \bibinfo{author}{Ye,
  P.}
\newblock \bibinfo{title}{All optical {XOR} logic gates: technologies and
  experiment demonstrations}.
\newblock \emph{\bibinfo{journal}{IEEE Commun. Mag.}}
  \textbf{\bibinfo{volume}{43}}, \bibinfo{pages}{S19--S24}
  (\bibinfo{year}{2005}).

\bibitem{ref:wang2012uif}
\bibinfo{author}{Wang, Y.-D.} \& \bibinfo{author}{Clerk, A.~A.}
\newblock \bibinfo{title}{Using interference for high fidelity quantum state
  transfer in optomechanics}.
\newblock \emph{\bibinfo{journal}{Phys. Rev. Lett.}}
  \textbf{\bibinfo{volume}{108}}, \bibinfo{pages}{153603}
  (\bibinfo{year}{2012}).

\bibitem{ref:buckle1986ai}
\bibinfo{author}{Buckle, S.~J.}, \bibinfo{author}{Barnett, S.~M.},
  \bibinfo{author}{Knight, P.~L.}, \bibinfo{author}{Lauder, M.~A.} \&
  \bibinfo{author}{Pegg, D.~T.}
\newblock \bibinfo{title}{Atomic interferometers}.
\newblock \emph{\bibinfo{journal}{Optica Acta: International Journal of
  Optics}} \textbf{\bibinfo{volume}{33}}, \bibinfo{pages}{1129--1140}
  (\bibinfo{year}{1986}).

\bibitem{ref:kosachiov1992cpm}
\bibinfo{author}{Kosachiov, D.~V.}, \bibinfo{author}{Matisov, B.~G.} \&
  \bibinfo{author}{Rozhdestvensky, Y.~V.}
\newblock \bibinfo{title}{Coherent phenomena in multilevel systems with closed
  interaction contour}.
\newblock \emph{\bibinfo{journal}{J. Phys. B}} \textbf{\bibinfo{volume}{25}},
  \bibinfo{pages}{2473--2488} (\bibinfo{year}{1992}).

\bibitem{ref:tian2012asc}
\bibinfo{author}{Tian, L.}
\newblock \bibinfo{title}{Adiabatic state conversion and pulse transmission in
  optomechanical systems}.
\newblock \emph{\bibinfo{journal}{Phys. Rev. Lett.}}
  \textbf{\bibinfo{volume}{108}}, \bibinfo{pages}{153604}
  (\bibinfo{year}{2012}).

\bibitem{ref:palomaki2013cst}
\bibinfo{author}{Palomaki, T.}, \bibinfo{author}{Harlow, J.},
  \bibinfo{author}{Teufel, J.}, \bibinfo{author}{Simmonds, R.} \&
  \bibinfo{author}{Lehnert, K.}
\newblock \bibinfo{title}{Coherent state transfer between itinerant microwave
  fields and a mechanical oscillator}.
\newblock \emph{\bibinfo{journal}{Nature}} \textbf{\bibinfo{volume}{495}},
  \bibinfo{pages}{210--214} (\bibinfo{year}{2013}).

\bibitem{ref:mitchell2016scd}
\bibinfo{author}{Mitchell, M.} \emph{et~al.}
\newblock \bibinfo{title}{Single-crystal diamond low-dissipation cavity
  optomechanics}.
\newblock \emph{\bibinfo{journal}{Optica}} \textbf{\bibinfo{volume}{3}},
  \bibinfo{pages}{963--970} (\bibinfo{year}{2016}).

\bibitem{ref:lake2018oit}
\bibinfo{author}{Lake, D.~P.}, \bibinfo{author}{Mitchell, M.},
  \bibinfo{author}{Kamaliddin, Y.} \& \bibinfo{author}{Barclay, P.~E.}
\newblock \bibinfo{title}{Optomechanically induced transparency and cooling in
  thermally stable diamond microcavities}.
\newblock \emph{\bibinfo{journal}{ACS Photonics}} \textbf{\bibinfo{volume}{5}},
  \bibinfo{pages}{782--787} (\bibinfo{year}{2018}).

\bibitem{ref:mitchell2019oaw}
\bibinfo{author}{Mitchell, M.}, \bibinfo{author}{Lake, D.~P.} \&
  \bibinfo{author}{Barclay, P.~E.}
\newblock \bibinfo{title}{Optomechanically amplified wavelength conversion in
  diamond microcavities}.
\newblock \emph{\bibinfo{journal}{Optica}} \textbf{\bibinfo{volume}{6}},
  \bibinfo{pages}{832--838} (\bibinfo{year}{2019}).

\bibitem{ref:higginbotham2018heo}
\bibinfo{author}{Higginbotham, A.} \emph{et~al.}
\newblock \bibinfo{title}{Harnessing electro-optic correlations in an efficient
  mechanical converter}.
\newblock \emph{\bibinfo{journal}{Nat. Phys.}} \textbf{\bibinfo{volume}{14}},
  \bibinfo{pages}{1038--1042} (\bibinfo{year}{2018}).

\bibitem{ref:reddy2018high}
\bibinfo{author}{Reddy, D.~V.} \& \bibinfo{author}{Raymer, M.~G.}
\newblock \bibinfo{title}{High-selectivity quantum pulse gating of photonic
  temporal modes using all-optical {R}amsey interferometry}.
\newblock \emph{\bibinfo{journal}{Optica}} \textbf{\bibinfo{volume}{5}},
  \bibinfo{pages}{423--428} (\bibinfo{year}{2018}).

\bibitem{ref:wu2016not}
\bibinfo{author}{Wu, M.} \emph{et~al.}
\newblock \bibinfo{title}{Nanocavity optomechanical torque magnetometry and
  radiofrequency susceptometry}.
\newblock \emph{\bibinfo{journal}{Nat. Nanotechnol.}}
  \textbf{\bibinfo{volume}{12}}, \bibinfo{pages}{127--131}
  (\bibinfo{year}{2016}).

\bibitem{ref:khanaliloo2015hqv}
\bibinfo{author}{Khanaliloo, B.}, \bibinfo{author}{Mitchell, M.},
  \bibinfo{author}{Hryciw, A.~C.} \& \bibinfo{author}{Barclay, P.~E.}
\newblock \bibinfo{title}{High-${Q}/{V}$ monolithic diamond microdisks
  fabricated with quasi-isotropic etching}.
\newblock \emph{\bibinfo{journal}{Nano Lett.}} \textbf{\bibinfo{volume}{15}},
  \bibinfo{pages}{5131--5136} (\bibinfo{year}{2015}).

\bibitem{ref:mitchell2018rqo}
\bibinfo{author}{Mitchell, M.}, \bibinfo{author}{Lake, D.~P.} \&
  \bibinfo{author}{Barclay, P.~E.}
\newblock \bibinfo{title}{Realizing ${Q}>$ 300 000 in diamond microdisks for
  optomechanics via etch optimization}.
\newblock \emph{\bibinfo{journal}{APL Photonics}} \textbf{\bibinfo{volume}{4}},
  \bibinfo{pages}{016101} (\bibinfo{year}{2019}).

\end{thebibliography}

\begin{thebibliography}{10}

\bibitem{ref:supp_khanaliloo2015hqv}
\bibinfo{author}{Khanaliloo, B.}, \bibinfo{author}{Mitchell, M.},
  \bibinfo{author}{Hryciw, A.~C.} \& \bibinfo{author}{Barclay, P.~E.}
\newblock \bibinfo{title}{High-${Q}/{V}$ monolithic diamond microdisks
  fabricated with quasi-isotropic etching}.
\newblock \emph{\bibinfo{journal}{Nano Lett.}} \textbf{\bibinfo{volume}{15}},
  \bibinfo{pages}{5131--5136} (\bibinfo{year}{2015}).

\bibitem{ref:supp_mitchell2018rqo}
\bibinfo{author}{Mitchell, M.}, \bibinfo{author}{Lake, D.~P.} \&
  \bibinfo{author}{Barclay, P.~E.}
\newblock \bibinfo{title}{Realizing ${Q}>$ 300 000 in diamond microdisks for
  optomechanics via etch optimization}.
\newblock \emph{\bibinfo{journal}{APL Photonics}} \textbf{\bibinfo{volume}{4}},
  \bibinfo{pages}{016101} (\bibinfo{year}{2019}).

\bibitem{ref:supp_lake2018oit}
\bibinfo{author}{Lake, D.~P.}, \bibinfo{author}{Mitchell, M.},
  \bibinfo{author}{Kamaliddin, Y.} \& \bibinfo{author}{Barclay, P.~E.}
\newblock \bibinfo{title}{Optomechanically induced transparency and cooling in
  thermally stable diamond microcavities}.
\newblock \emph{\bibinfo{journal}{ACS Photonics}} \textbf{\bibinfo{volume}{5}},
  \bibinfo{pages}{782--787} (\bibinfo{year}{2018}).

\bibitem{ref:supp_aharonovich2011dp}
\bibinfo{author}{Aharonovich, I.}, \bibinfo{author}{Greentree, A.~D.} \&
  \bibinfo{author}{Prawer, S.}
\newblock \bibinfo{title}{Diamond photonics}.
\newblock \emph{\bibinfo{journal}{Nat. Photonics}}
  \textbf{\bibinfo{volume}{5}}, \bibinfo{pages}{397--405}
  (\bibinfo{year}{2011}).

\bibitem{ref:supp_mitchell2016scd}
\bibinfo{author}{Mitchell, M.} \emph{et~al.}
\newblock \bibinfo{title}{Single-crystal diamond low-dissipation cavity
  optomechanics}.
\newblock \emph{\bibinfo{journal}{Optica}} \textbf{\bibinfo{volume}{3}},
  \bibinfo{pages}{963--970} (\bibinfo{year}{2016}).

\bibitem{ref:supp_kippenberg2005arp}
\bibinfo{author}{Kippenberg, T.~J.}, \bibinfo{author}{Rokhsari, H.},
  \bibinfo{author}{Carmon, T.}, \bibinfo{author}{Scherer, A.} \&
  \bibinfo{author}{Vahala, K.~J.}
\newblock \bibinfo{title}{Analysis of radiation-pressure induced mechanical
  oscillation of an optical microcavity}.
\newblock \emph{\bibinfo{journal}{Phys. Rev. Lett.}}
  \textbf{\bibinfo{volume}{95}}, \bibinfo{pages}{033901}
  (\bibinfo{year}{2005}).

\bibitem{ref:supp_carmon2005temporal}
\bibinfo{author}{Carmon, T.}, \bibinfo{author}{Rokhsari, H.},
  \bibinfo{author}{Yang, L.}, \bibinfo{author}{Kippenberg, T.~J.} \&
  \bibinfo{author}{Vahala, K.~J.}
\newblock \bibinfo{title}{Temporal behavior of radiation-pressure-induced
  vibrations of an optical microcavity phonon mode}.
\newblock \emph{\bibinfo{journal}{Phys. Rev. Lett.}}
  \textbf{\bibinfo{volume}{94}}, \bibinfo{pages}{223902}
  (\bibinfo{year}{2005}).

\bibitem{ref:supp_rokhsari2005radiation}
\bibinfo{author}{Rokhsari, H.}, \bibinfo{author}{Kippenberg, T.~J.},
  \bibinfo{author}{Carmon, T.} \& \bibinfo{author}{Vahala, K.~J.}
\newblock \bibinfo{title}{Radiation-pressure-driven micro-mechanical
  oscillator}.
\newblock \emph{\bibinfo{journal}{Opt. Express}} \textbf{\bibinfo{volume}{13}},
  \bibinfo{pages}{5293--5301} (\bibinfo{year}{2005}).

\bibitem{ref:supp_kippenberg2002mct}
\bibinfo{author}{Kippenberg, T.~J.}, \bibinfo{author}{Spillane, S.~M.} \&
  \bibinfo{author}{Vahala, K.~J.}
\newblock \bibinfo{title}{Modal coupling in traveling-wave resonators}.
\newblock \emph{\bibinfo{journal}{Opt. Lett.}} \textbf{\bibinfo{volume}{27}},
  \bibinfo{pages}{1669--1671} (\bibinfo{year}{2002}).

\bibitem{ref:supp_yan2003mot}
\bibinfo{author}{Yan, L.-S.}, \bibinfo{author}{Yu, Q.},
  \bibinfo{author}{Willner, A.~E.} \& \bibinfo{author}{Shi, Y.}
\newblock \bibinfo{title}{Measurement of the chirp parameter of electro-optic
  modulators by comparison of the phase between two sidebands}.
\newblock \emph{\bibinfo{journal}{Opt. Lett.}} \textbf{\bibinfo{volume}{28}},
  \bibinfo{pages}{1114--1116} (\bibinfo{year}{2003}).

\bibitem{ref:supp_dong2012odm}
\bibinfo{author}{Dong, C.}, \bibinfo{author}{Fiore, V.},
  \bibinfo{author}{Kuzyk, M.~C.} \& \bibinfo{author}{Wang, H.}
\newblock \bibinfo{title}{Optomechanical dark mode}.
\newblock \emph{\bibinfo{journal}{Science}} \textbf{\bibinfo{volume}{338}},
  \bibinfo{pages}{1609--1613} (\bibinfo{year}{2012}).

  
\end{thebibliography}
\end{document}